\documentclass[fleqn,usenatbib]{mnras}

\usepackage{newtxtext,newtxmath}

\usepackage[T1]{fontenc}
\usepackage{ae,aecompl}

\usepackage{graphicx}	
\usepackage{amsmath}	

\usepackage{amssymb}	

\usepackage[utf8]{inputenc} 
\usepackage[T1]{fontenc}    
\usepackage{hyperref}       
\usepackage{url}            
\usepackage{booktabs}       
\usepackage{amsfonts}       
\usepackage{nicefrac}       
\usepackage{microtype}      
\usepackage{graphicx}       
\usepackage{float}          
\usepackage{gensymb}
\usepackage{cleveref}
\usepackage{textcomp}

\usepackage{scalerel,tikz}
\usetikzlibrary{svg.path}
\definecolor{orcidlogocol}{HTML}{A6CE39}
\tikzset{orcidlogo/.pic={
 \fill[orcidlogocol] svg{M256,128c0,70.7-57.3,128-128,128C57.3,256,0,198.7,0,128C0,57.3,57.3,0,128,0C198.7,0,256,57.3,256,128z};
 \fill[white] svg{M86.3,186.2H70.9V79.1h15.4v48.4V186.2z}
 svg{M108.9,79.1h41.6c39.6,0,57,28.3,57,53.6c0,27.5-21.5,53.6-56.8,53.6h-41.8V79.1z M124.3,172.4h24.5c34.9,0,42.9-26.5,42.9-39.7c0-21.5-13.7-39.7-43.7-39.7h-23.7V172.4z}
 svg{M88.7,56.8c0,5.5-4.5,10.1-10.1,10.1c-5.6,0-10.1-4.6-10.1-10.1c0-5.6,4.5-10.1,10.1-10.1C84.2,46.7,88.7,51.3,88.7,56.8z};
}}
\newcommand\orcidicon[1]{\href{https://orcid.org/#1}{\mbox{\scalerel*{
\begin{tikzpicture}[yscale=-1,transform shape]
\pic{orcidlogo};
\end{tikzpicture}
}{|}}}}

\crefformat{section}{\S#2#1#3} 
\crefformat{subsection}{\S#2#1#3}
\crefformat{subsubsection}{\S#2#1#3}
\crefformat{table}{Table #2#1#3}
\crefformat{figure}{Figure #2#1#3}
\crefformat{equation}{Eq. #2#1#3}

\def\aj{AJ}                   
\def\apj{ApJ}  
\def\aap{A\&A}
\def\pasa{PASA}
\def\araa{Annual Review of A\&A}
\def\mnras{MNRAS}

\makeatletter
\newcommand\notsotiny{\@setfontsize\notsotiny\@vipt\@viipt}
\makeatother

\title[A Radio Study of Magnetic Fields in the SMC]{A Radio Polarisation Study of Magnetic Fields in the Small Magellanic Cloud}

\author[J. D. Livingston et al.]{
J. D. Livingston,$^{\orcidicon{0000-0002-4090-8000}\,1}$\thanks{E-mail: jack.david.livingston+academic@gmail.com}
N. M. McClure-Griffiths,$^{\orcidicon{0000-0003-2730-957X}\,1}$
S. A. Mao,$^{\orcidicon{000-0001-8906-7866}\,2}$
Y. K. Ma,$^{\orcidicon{0000-0003-0742-2006}\,1,2}$
\newauthor
B. M. Gaensler,$^{\orcidicon{0000-0002-3382-9558}\,3}$
G. Heald,$^{\orcidicon{0000-0002-2155-6054}\,4}$
and A. Seta$^{\orcidicon{0000-0001-9708-0286}\,1}$
\\
$^{1}$Research School of Astronomy \& Astrophysics, The Australian National University, Canberra ACT 2611, Australia\\
$^{2}$ Max-Planck-Institut für Radioastronomie, 
Auf dem Hügel 69, 53121 Bonn, Germany\\
$^{3}$Dunlap Institute for Astronomy and Astrophysics, 
University of Toronto, Toronto ON M5S 3H4, Canada\\
$^{4}$CSIRO, Space and Astronomy, PO Box 1130, Bentley, WA 6102, Australia}

\date{Accepted XXX. Received YYY; in original form ZZZ}

\pubyear{2021}

\begin{document}
\label{firstpage}
\pagerange{\pageref{firstpage}--\pageref{lastpage}}
\maketitle

\begin{abstract}
Observing the magnetic fields of low-mass interacting galaxies tells us how they have evolved over cosmic time and their importance in galaxy evolution. We have measured the Faraday rotation of 80 extra-galactic radio sources behind the Small Magellanic Cloud (SMC) using the CSIRO Australia Telescope Compact Array (ATCA) with a frequency range of 1.4 -- 3.0 GHz. Both the sensitivity of our observations and the source density are an order of magnitude improvement on previous Faraday rotation measurements of this galaxy. The SMC generally produces negative rotation measures (RMs) after accounting for the Milky Way foreground contribution, indicating that it has a mean coherent line-of-sight magnetic field strength of $-0.3\pm0.1\mu$G, consistent with previous findings. We detect signatures of magnetic fields extending from the north and south of the Bar of the SMC. The random component of the SMC magnetic field has a strength of $\sim 5\mu$G with a characteristic size-scale of magneto-ionic turbulence $< 250$ pc, making the SMC like other low-mass interacting galaxies. The magnetic fields of the SMC and Magellanic Bridge appear similar in direction and strength, hinting at a connection between the two fields as part of the hypothesised `pan-Magellanic' magnetic field. 
\end{abstract}


\section{Introduction}
\label{sec:introSMC}

With energy densities comparable to thermal gas and cosmic rays \citep{Heiles2012}, magnetic fields in star-forming galaxies play an important role in the dynamics of gas flows and in the turbulent interstellar medium (ISM) \citep{Beck2013}. Low mass interacting galaxies like the Small and Large Magellanic Clouds (SMC and LMC) make up the bulk of the population of galaxies in the early universe \citep{Tolstoy2009,Boylan-Kolchin2015}. Observing the magnetic fields of low-mass interacting galaxies tells us about how they have evolved over cosmic time and their importance in galaxy evolution. We know that areas of tidal interaction in interacting galaxies can have strong magnetic fields \citep{Chyzy2004,Basu2017}. The random components of these fields are typically stronger than the coherent components, which reduces the regularity of the field \citep{Chyzy2004,Basu2017}. Interacting galaxies and low-mass galaxies have also been shown to magnetise their surroundings \citep{Chyzy2011,Drzazga2011}. 

\begin{figure}
    \centering
    \includegraphics[width=\columnwidth]{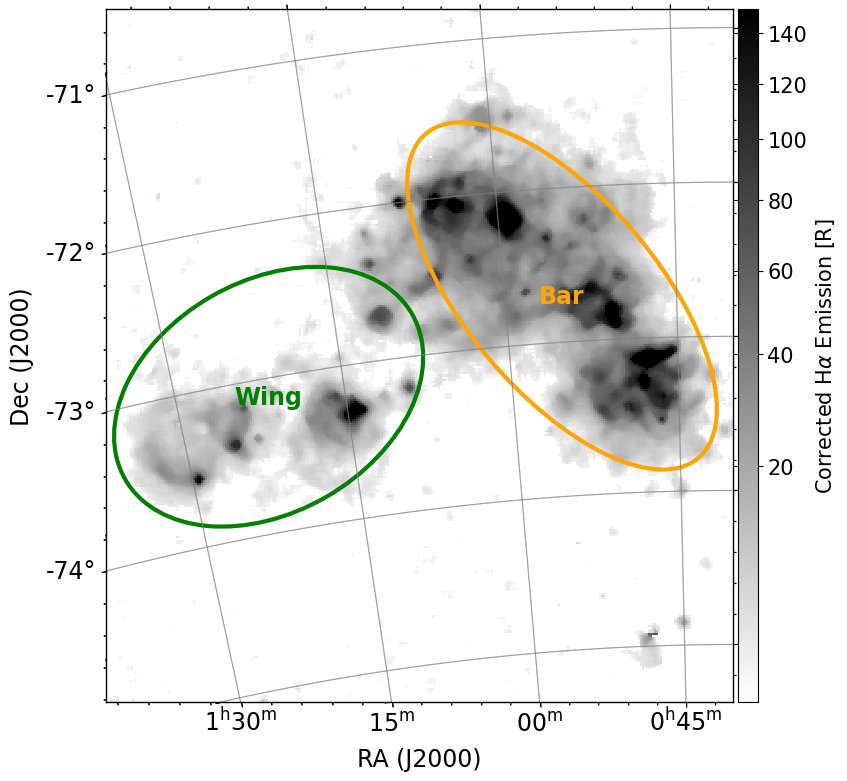}
    \caption{Image of the SMC from the \protect\cite{Gaustad2001} H$\alpha$ map. The orange ellipse covers the Bar region and the green ellipse covers the Wing region of the SMC \protect\citep{Bergh2007}.}
    \label{fig:SMC}
\end{figure}

In this paper, we aim to investigate the strength and structure of the line-of-sight (LOS) magnetic field of the low-mass interacting galaxy, the SMC, and its surroundings. The SMC is at a distance of 63 $\pm 1 \,\mathrm{kpc}$ from Earth \citep{Cioni2000}. Given its proximity to the Milky Way (MW) and to the LMC, it is subjected to significant tidal force \citep{Besla2012}. The two Clouds are likely on their first or second pass of the MW \citep{Besla2007}. Between the LMC and SMC is an inter-cloud region called the Magellanic Bridge (MB) that is made up of gas that has been tidally stripped/shared from the SMC and brought towards the LMC \citep{Bergh2007,Besla2010,Besla2012}. The SMC is split into two main regions of star formation, the Bar and Wing (the locations of these regions are shown in Figure \ref{fig:SMC}). The Bar of the SMC contains the majority of the SMC's star formation and mass \citep{Bergh2007}. The Wing of the SMC connects the SMC to the MB and is controlled by tidal forces between the LMC and SMC \citep{Bergh2007}.


Observing how much the polarisation angle of linearly polarised radiation rotates in magnetised plasma via the Faraday effect is a sensitive method for measuring magnetic fields. This radiation comes from distant radio galaxies, extra-galactic background sources (EGSs), or synchrotron-emitting sources such as supernova remnants or diffuse ISM within the target galaxy itself. At a wavelength, $\lambda$, the polarisation angle, $\chi$, is determined using the Stokes parameters \textit{Q} and \textit{U},
\begin{equation}
\label{eqn:chi}
    \chi (\lambda) = \frac{1}{2} \tan^{-1} \left[\frac{U(\lambda)}{Q(\lambda)}\right].
\end{equation}
We define the fractional polarisation signal in Stokes \textit{Q}, \textit{U}, and Stokes \textit{V} as $q = Q/I$, $u = U/I$, and $v = V/I$, where \textit{I} is the measured Stokes \textit{I}. The total linear polarised intensity, \textit{P}, and total linear fraction of polarisation \textit{p} are defined as $P^2 = Q^2 + U^2$ and $p^2 = q^2 + u^2$. The variation of $\chi$ with $\lambda^2$ measures the Faraday rotation of the foreground medium or Faraday screen. This is called Rotation Measure (RM). RM relates to $\chi$ as, $ \chi = \chi_0 + \mathrm{RM} \lambda^{2}$, where $\chi_0$ is the intrinsic polarisation angle of a source. With a collection of RMs from background sources, we can form an RM Grid that can map the RM signature of a galaxy. This relates to the LOS thermal electron density, $n_e$ (measured in $\mathrm{cm}^{-3}$) and the magnetic field component along the line of sight, $B_{||}$ (measured in micro-Gauss), as,
\begin{equation}
\mathrm{RM} \equiv\mathscr{C}\int_{r=0}^{r=d} n_e\,B_{||} \cdot d\,\textbf{\textit{r}}\, (\mathrm{rad}\,\mathrm{m}^{-2}). \label{eqn:RM}
\end{equation}
Here, $d$ is the distance to the RM source in pc, \textit{d}\textbf{\textit{r}} is the incremental displacement along the LOS measured in pc from the observer to the source, and $\mathscr{C}$ is a conversion constant, $\mathscr{C} = 0.812 \, \mathrm{rad}\,\mathrm{m}^{-2}\,\mathrm{pc}^{-1}\, \mathrm{cm}^{3}\,\mu \mathrm{G}^{-1}$. We adopt the vector orientation and sign convention for Faraday rotation as outlined by \cite{Ferriere2021}.

The relation in \cref{eqn:RM} reduces the observed Faraday rotation to a single measurement where there is no emission of polarised radiation along the LOS. Thus, it may not reflect the actual nature of the magneto-ionic environment along the LOS. By contrast, Faraday depth ($\phi$) \citep{Burn1966} measures the Faraday rotation as a function of position along the LOS,
\begin{equation}
    \phi (r)=\mathscr{C}\int_{0}^{r} n_e\,B_{||} \cdot \mathrm{d}\,\textbf{\textit{r'}}\,(\mathrm{rad}\,\mathrm{m}^{-2}).
    \label{eqn:phi}
\end{equation}
We note that $r$ is at any distance along the path, differentiating it from \cref{eqn:RM}. Previous studies have found that more than half of EGSs are modelled accurately with multiple RM components and other Faraday effects \citep{Burn1966,Anderson2015,OSullivan2017,Livingston2021}. Thus, we require a way of measuring these multiple RM components. 

One way of potentially resolving these multiple RM components is RM synthesis. \cite{Burn1966} introduced a relation between the complex polarised surface brightness of a source, $\mathcal{P}$, and the distribution of Faraday depths. This is called a Faraday dispersion function ($F(\phi)$) or `Faraday spectrum'. \cite{Brentjens2005} reformulated this relation for use with modern broadband radio telescopes. $\mathcal{P}$ relates to $F(\phi)$ as,
\begin{equation}
    \mathcal{P}(\lambda^{2}) = \frac{1}{\pi} \int_{-\infty}^{\infty} F(\phi) e^{2i \phi \lambda^{2}} \mathrm{d}\phi.
\end{equation}
The use of the inverse relation to determine $F(\phi$) from $\mathcal{P}(\lambda^{2})$ is highly dependent on the coverage of $\lambda^{2}$. The $F(\phi$) of a single thin Faraday screen under infinite sampling would be,
\begin{equation}
    F(\phi) \approx \mathcal{P}_0 \delta (x - \alpha),
\end{equation}
where $\alpha$ is the centre of a Dirac delta function, $\delta (x - \alpha)$, measured in Faraday depth and $\mathcal{P}_0$ is the amplitude of the Faraday depth peak. However, in reality this delta function is convolved with a spread function due to incomplete wavelength sampling. This spread function is called the Rotation Measure Spread Function (RMSF). The range of observed $\lambda^2$ determine the width of the RMSF, the channel widths of the observed $\lambda^2$ determine the maximum measurable Faraday depth, and the minimum value for the observed $\lambda^2$ determines the maximum measurable width in Faraday depth of a Faraday depth component. These are quantified by \cite{Brentjens2005,Schnitzeler2009,Dickey2019}. We have used RM synthesis to determine $F(\phi)$ as it allows us to resolve different Faraday depth components that may occur along the LOS.

The first studies of the magnetic field of the SMC were done using the star-light polarisation of 147 SMC stars from \cite{Mathewson1970,Mathewson1970b,Schmidt1970,Deinzer1973,Schmidt1976,Wayte1990,Magalhaes1990}. These studies found a plane of sky magnetic field for the SMC aligned with the Magellanic Bridge which served as the basis of the `pan-Magellanic' magnetic field hypothesis. Radio continuum images of the SMC at 1.4, 2.3, 2.5, 4.8, and 8.6 GHz \citep{Loiseau1987,Haynes1990} have been used to estimate the total plane of sky magnetic field of the SMC assuming energy equipartition. The findings suggested that the SMC had a large scale magnetic field \citep{Haynes1990} with a total field strength of $\sim 5 \mu$G \citep{Loiseau1987}. While the energy equipartition assumption for the SMC is inconsistent with the findings of \cite{Chi1993}, it may in fact be consistent if the energy of cosmic ray electrons are taken into account \citep{Pohl1993,Mao2008}.

\cite{Mao2008} conducted the first RM Grid experiment using background EGSs to study the LOS magnetic field of the SMC. They found 10 sight-lines that passed through the SMC from which RM determinations were possible. After the subtraction of the MW foreground RM contribution, 9 of the sources had a negative RM, indicating a coherent magnetic field for the SMC. They found a coherent component of $-0.19 \pm 0.06 \mu$G directed uniformly away from us and that the SMC has a random magnetic field component that is much stronger than the coherent with an average strength of $1.9\mu$G \citep{Mao2008}. As random magnetic field strengths are enhanced strongly by major merger events \citep{Basu2017}, we expect the random magnetic field of the SMC to be enhanced. Additionally, for local low-mass galaxies, star-formation magnetises their surroundings up to a field strength of $0.1 \mu$G within 5 kpc \citep{Chyzy2011}.

Physically, the formation of galactic magnetic fields are normally explained by a dynamo theory. On large-scales, a dynamo called the $\alpha - \omega$ dynamo typically generates the coherent component of the magnetic field of a galaxy. On small-scales, magnetic field formation is typically done by a turbulent dynamo that generates and modulates the random magnetic field component of galaxies \citep{Brandenburg2005}. For the SMC, \cite{Mao2008} also found that the strength of the coherent field was incompatible with the standard $\alpha - \omega$ dynamo. This result has also been found for other low-mass galaxies \citep{Chyzy2011,Jurusik2014}. The small-scale dynamo depends only on turbulence \citep{Seta2020} and thus is expected to be active in the SMC. 

In this paper, we aim to study the LOS magnetic field of the SMC using broadband Faraday rotation measurements. The observations that constitute the data of this paper are five times more sensitive than previous measurements. This allows us to improve the number of sight-lines whose RMs can be sampled for the RM Grid of the SMC by an order of magnitude. We present the calculated RMs of 80 sources towards the SMC with RM synthesis in Section \ref{sec:results} along with our discussion of trends in the spatial distributions of the data and comparisons to previous studies of the magnetism of the SMC. In Section \ref{sec:discuss}, we provide an estimation of the LOS magnetic field of the SMC from RM and discuss the magnetic field of the SMC. We give our conclusions and suggestions for future work in Section \ref{sec:conclude}.

\section{Methods}
\label{sec:data}
\subsection{Observations and Data Reduction}
The observations for this study were obtained with the Australia Telescope Compact Array (ATCA) for \cite{Jameson2019}\footnote{Obtained from the \href{https://atoa.atnf.csiro.au/}{Australia Telescope Online Archive}}. The primary purpose of the project was to observe the $\ion{H}{i}$ absorption of the SMC. The fields were chosen based on the presence of strong sources with potentially high $\ion{H}{i}$ absorption. We reprocessed the \cite{Jameson2019} data set for polarisation analysis across a frequency range of 1.1 -- 3.1 GHz, divided among 2048 channels. Observations were conducted with two different array configurations 6A and 6C. 6A has a minimum baseline of 336.7 m and a maximum of 5938.8 m, 6C has a minimum baseline of 153.1 m and a maximum of 6000 m.

\begin{figure}
    \centering
    \includegraphics[width=\columnwidth]{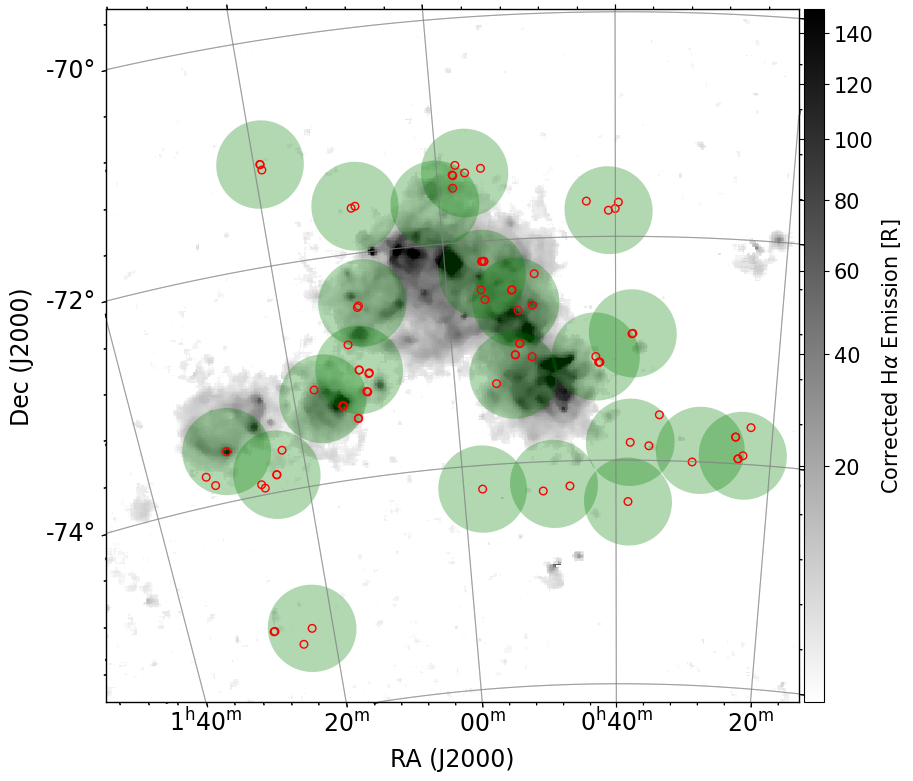}
    \caption{Distribution of fields on top of a background image of the \protect\cite{Gaustad2001} H$\alpha$ map. The green regions mark the position of each field, and the red circles mark the position of each polarised source.}
    \label{fig:field_map}
\end{figure}

We have 22 fields covering selected regions of the SMC as shown in \cref{fig:field_map}, each with an average maximum size of 1 deg$^{2}$. The average telescope resolving element (synthesised beam) size for each field was $(6 \times 5)$ arcsec$^{2}$ calculated from the highest frequency observed. Each field was observed for 11 hours in total between 2016 and 2018. The primary flux density and bandpass calibrator for all fields was PKS B1934--638. The linearly polarised fraction of PKS B1934--638 is known to be < 0.2\% \citep{Rayner2000}. The polarised leakages were determined and corrected using the parallactic angle coverage of the phase calibrator, PKS B0252--712. 

For data reduction we used the software package {\sc Miriad} \citep{Sault1995}. We completed data flagging by hand in {\sc Miriad} using the sum-threshold method. After flagging, we applied the bandpass and polarised leakage solutions to the data and created maps of Stokes parameters \textit{I}, \textit{Q}, \textit{U}, and \textit{V} using {\sc invert} and {\sc clean} tasks. The robust visibility weighting was set to +0.8, with $1''$ per pixel. We binned these data in groups of 30 MHz for the mapping process. This was done to increase the signal-to-noise ratio of channel maps. The RM-synthesis capabilities for this study are shown in Table \ref{tab:faracap}.

\begin{table}
    \centering
    \normalsize
    \setlength{\tabcolsep}{0.5em}
    \begin{tabular}{c c}
    \hline
        Parameter & Value \\
        \hline
         Mean Frequency Range (1) (GHz) & 1.4 -- 3.0 \\
         Mean Channel Width (2) (MHz) & 33.8 \\
         Mean Sensitivity (3) (mJy/beam) & 0.2 \\
         Mean Synthesised Beam (4) (arcsec$^{2}$) & $6\times5$ \\
         \hline
         Mean FWHM of RMSF, $\delta \phi$ (5) ($\mathrm{rad}\,\mathrm{m}^{-2}$) & 176 \\
         Mean max-scale in $\phi$ (6) ($\mathrm{rad}\,\mathrm{m}^{-2}$) & 316 \\
         Mean maximum measurable $\phi$, $|\phi_{\mathrm{max}}|$ (7) ($\mathrm{rad}\,\mathrm{m}^{-2}$) & 3139\\
         \hline
    \end{tabular}
    \caption{Table of observational and RM-synthesis capabilities based on frequency range and channel size. Row 3 is the mean Stokes V noise over all fields, row 4 is the mean synthesised beam dimensions. The quantities of rows 5, 6, and 7 are outlined in \protect\cite{Brentjens2005,Dickey2019}. The frequency range is smaller than the total observed of 1.1 -- 3.1 GHz due to flagged RFI. The average channel width is larger than the width used to produce the image cubes due to flagging of RFI.}
    \label{tab:faracap}
\end{table}

\subsubsection{Source Finding}

We conducted source finding using the BANE and Aegean applications \citep{Hancock2012,Hancock2018} on each total Stokes \textit{I} intensity Multi Frequency Synthesis (MFS) image. We extracted Stokes data by collecting the Stokes \textit{Q} and \textit{U} values at the Stokes \textit{I} peak position. The Stokes \textit{V} cube was used to find the RMS of the Stokes \textit{Q} and \textit{U} cubes, taking 41 x 41 pixel boxes centred on each source position. A source that had a mean \textit{P} signal-to-noise ratio lower than six averaged over all channels or had fewer than 50\% of its original channels remaining after flagging was considered not polarised. The location and sizes of the 80 polarised sources found are shown in Table \ref{tab:observed} and the spatial distribution of sources is shown in \cref{fig:field_map}.

\begin{table*}
\tiny
\centering
\setlength{\tabcolsep}{0.85em}
    \begin{tabular}{c c c c c c c c c c c c c c}
    \hline
    Source (1) & RAJ2000 (2) & Error (3) & DECJ2000 (4) & Error (5) & $\mathrm{s_{maj \times min}}$ (6) & $S_{1.4\,\mathrm{Stokes}\,I}$ (7) & $S_{1.4\,p}$ (8) & $\mathrm{RM}_{\textrm{ raw}}$ (9) & Error (10) & $\mathrm{RM}_{\textrm{MGS, raw}}$ (11) &  Error (12)  & $\mathrm{RM_{SMC}}$ (13) & Error (14) \\
    Name & (h m s) & (s) & ($^{\circ}$ $\,'$ $\,''$)  & ($''$) & (arcsec$^{2}$) & (mJy) & (\%) & ($\textrm{rad}\,\textrm{m}^{-2}$) & ($\pm$) & ($\textrm{rad}\,\textrm{m}^{-2}$) & ($\pm$)
 & ($\textrm{rad}\,\textrm{m}^{-2}$) & ($\pm$) \\
    \hline
    \hline
J002248.1-734008.1 & 00:22:48.1 & 0.1 & -73:40:08.1 & 0.5 & 5.6 $\times$ 5.5 & $25.6\pm0.2$&$ 4.3\pm0.8$ & +36.2 & 11.5 &  &  & -2.0  & 15.8\\
J002335.2-735529.1 & 00:23:35.2 & 0.1 & -73:55:29.1 & 0.5 & 6.7 $\times$ 4.9 & $46.4\pm0.6$&$ 4.3\pm1.2$ & +54.3 & 5.0 &  &  & +16.2  & 12.0\\
MGS2008 42 & 00:24:11.9&0.1 & -73:57:17.9&0.5 & 7.0 $\times$ 4.8 & $66.7\pm0.4$&$ 6.4\pm0.6$ & +51.4 & 4.5 & +54.0 & 8.0 & +13.5 & 11.8 \\

\hline
    \end{tabular}
    \caption{Example table of observed sources; a full machine-readable table is available as part of the supplementary material provided online. The columns are: Source Name, RAJ2000 and DECJ2000, approximate major and minor axes, Stokes \textit{I} intensity and fractional polarisation (\textit{p}) at 1.4 GHz, our RMs before MW foreground subtraction, previous measurements of the sources from \protect\cite{Mao2008} where they exist, and our RM after MW foreground subtraction in units of $\mathrm{rad}\,\mathrm{m}^{-2}$. Column 14 errors are the combined-in-quadrature errors from the measurement of RM, $\sigma_{\mathrm{RM,EG}}$ and $\sigma_{\mathrm{RM,MW}}$, and the MW foreground. Sources with an asterisk were determined by \protect\cite{Mao2008} to lie behind the SMC.}
    \label{tab:observed}
\end{table*}

\subsubsection{Faraday Depth Spectra}

To compute the Faraday depth spectrum for each source we used the {\sc rm synthesis} \citep{Brentjens2005} and {\sc rm clean} \citep{Heald2009,Heald2017} algorithms from the Canadian Initiative for Radio Astronomy Data Analysis (CIRADA) tool-set \href{https://github.com/CIRADA-Tools/RM-Tools/tree/v1.0.1}{RMtools 1D v1.0.1} \citep{Purcell2020}. This was done on the fractional polarisation data for each source with a $\phi$ range of $\pm 3000\,\mathrm{rad\,m^{-2}}$, which is informed by the RM-synthesis capabilities shown in \cref{tab:faracap}. As we will show in \cref{sec:complexity} most of our target EGs are Faraday simple and therefore we report the peak Faraday depth within each spectrum as the RM of the source. The RM for each source is shown in \cref{tab:observed}. For {\sc rm clean}, the cleaning limit was set to three times the noise in Stokes \textit{q} and \textit{u}. An example of a cleaned Faraday spectrum is shown in \cref{fig:exampleFDF}. All associated $F(\phi)$ spectra and Stokes information are shown in the supplementary material provided online. 

\subsubsection{Off-axis Polarisation Leakage}
\label{sec:offaxis}
As some of our sources are near the edges of fields, off-axis polarisation leakage may affect the determination of RM. \cite{Eyles2020} found that for ATCA observations, sources separated from the centre of the beam by more than 2/3 of the primary beam FWHM showed significant linear polarisation leakage, up to levels of 1.4\%. For all our sources that were further than 2/3 of the primary beam FWHM away from the centre of a field, we tested how off-axis leakage affected observed RM using the same method as used by \cite{Ma2019}. We set the leakage amplitude at the predicted leakage percentage from \cite{Eyles2020} for each source and found the RM of the new data using {\sc rm synthesis} and {\sc rm clean}. We repeated this process 1000 times for each source, choosing a new constant leakage polarisation angle for each iteration. There was no change to RM across all sources. This is expected as adding a leakage polarisation vector to $\mathcal{P}$ with a constant polarisation angle and low overall polarisation will create a low amplitude Faraday depth of $0\,\mathrm{rad\,m^{-2}}$ within each $F(\phi)$ spectrum, which is at a lower amplitude than the true RM. However, the addition of a leakage Faraday depth within $F(\phi)$ will increase the measured Faraday complexity of a source. As such, we have not included the 24 sources that were further than 2/3 of the primary beam FWHM away from the centre of a field in the analysis of Faraday complexity in \cref{sec:complexity}. 

\begin{figure*}
    \centering
    \includegraphics[width=1.875\columnwidth]{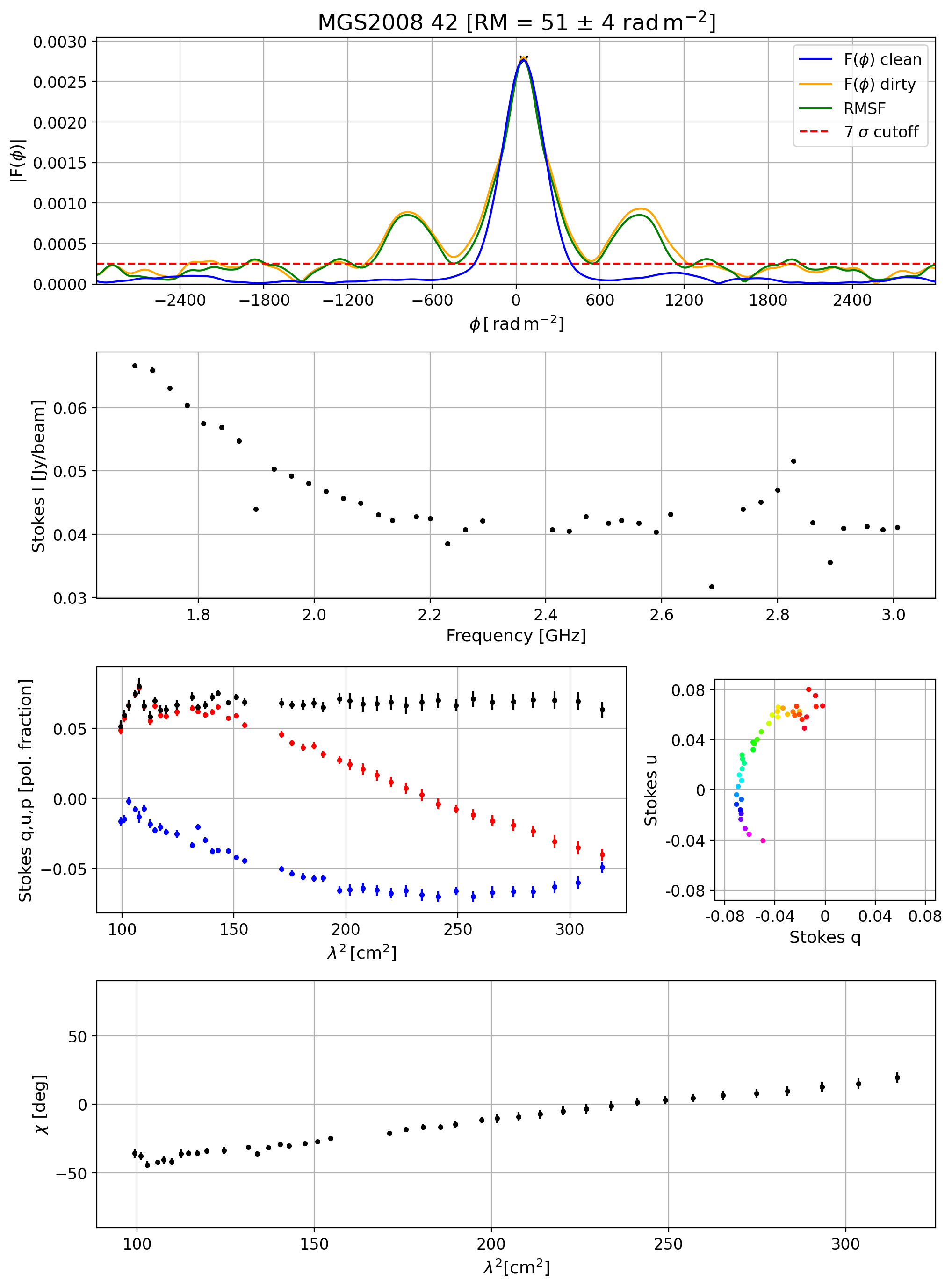}
    \caption{Example $F(\phi)$ for MGS2008 42, the full set can be found in the supplementary material provided online. In order from left to right, top to bottom; \emph{Panel 1:} The dirty $F(\phi)$ is shown in orange, the clean $F(\phi)$ is shown in blue, the RMSF is shown in green, the peak cutoff line (of seven times the noise) is shown in red, and the black crosses indicate peaks within the spectrum. \emph{Panel 2:} Stokes \textit{I} against frequency. \emph{Panel 3:} Fractional Stokes \textit{q} (blue), \textit{u} (red), and fractional polarised intensity, \textit{p}, (black) against $\lambda^{2}$. These fractional parameters are defined in \cref{sec:introSMC}. \emph{Panel 4:} Stokes \textit{q} against \textit{u}, coloured based on $\lambda^{2}$ ordered from larger $\lambda^{2}$ in red to smaller $\lambda^{2}$ in pink. \emph{Panel 5:} Polarisation angle, $\chi$, against $\lambda^{2}$.}
    \label{fig:exampleFDF}
\end{figure*}

\subsection{MW Foreground Subtraction}
\label{sec:foreground}
An RM measurement through the SMC will have many contributing factors,
\begin{equation}
    \mathrm{RM_{raw}} = \mathrm{RM_{int}} + \mathrm{RM_{IGM}} + \mathrm{RM_{SMC}} +\mathrm{RM_{MW}}.
\end{equation}
Here $\mathrm{RM_{raw}}$ is the observed RM, $\mathrm{RM_{int}}$ is the contribution from the sources themselves, $\mathrm{RM_{IGM}}$ is the contribution from the intergalactic medium, $\mathrm{RM_{SMC}}$ is the contribution from the SMC, and $\mathrm{RM_{MW}}$ is the contribution from the Milky Way. Typically, the magnitude of $\mathrm{RM_{IGM}} + \mathrm{RM_{int}}$ is between 1 -- 10 $\,\mathrm{rad\,m^{-2}}$ \citep{OSullivan2017}. Although the contribution of $\mathrm{RM_{IGM}} + \mathrm{RM_{int}}$ to the total RM is negligible as compared to $\mathrm{RM_{MW}}$ and $\mathrm{RM_{SMC}}$, they contribute to the uncertainty of RM. \cite{Schnitzeler2010} calculated the extragalactic contribution to the scatter of RMs, which includes both the scatter from $\mathrm{RM_{IGM}}$ and $\mathrm{RM_{int}}$, as $\sigma_{\mathrm{RM,EG}} \approx 6\, \mathrm{rad\,m^{-2}}$. We have included this scatter as part of the uncertainty of $\mathrm{RM_{SMC}}$ in \cref{tab:observed}.

To determine the Faraday depth associated with the SMC, we consider the LOS MW foreground model from \cite{Mao2008}. The \cite{Mao2008} model used 60 extra-galactic sources close to the SMC to estimate $\phi_{\textrm{MW}}$. To estimate this foreground, we used the following equation,
    \begin{equation}
        \phi_{\textrm{MW}} = (+46.1 \pm 4.1) - (+4.9 \pm 0.9) \times a\,\textrm{rad}\,\textrm{m}^{-2}.
        \label{eqn:fore}
    \end{equation}
Here $a = \mathrm{RA}\times\cos(\mathrm{DEC})$, where RA and DEC are in degrees of arc. The data were insufficient to create a foreground declination (DEC) function. The MW foreground model that we use only accounts for the smooth contribution of the MW. To account for the random component of RM from the MW we use $\sigma_{\mathrm{RM,MW}} \approx 8\, \mathrm{rad\,m^{-2}}$ from \cite{Schnitzeler2010} and include this as part of the uncertainty of $\mathrm{RM_{SMC}}$ in \cref{tab:observed}. We did not use the Galactic Faraday rotation sky 2020 model from \cite{Hutschenreuter2021} as an estimation of the contribution of the MW on $\mathrm{RM_{raw}}$. This model includes sources that \cite{Mao2008} determined to lie behind the SMC and as such cannot separate the contribution of $\mathrm{RM_{MW}}$ and $\mathrm{RM_{SMC}}$. 

\begin{figure*}
    \centering
    \includegraphics[width=2\columnwidth]{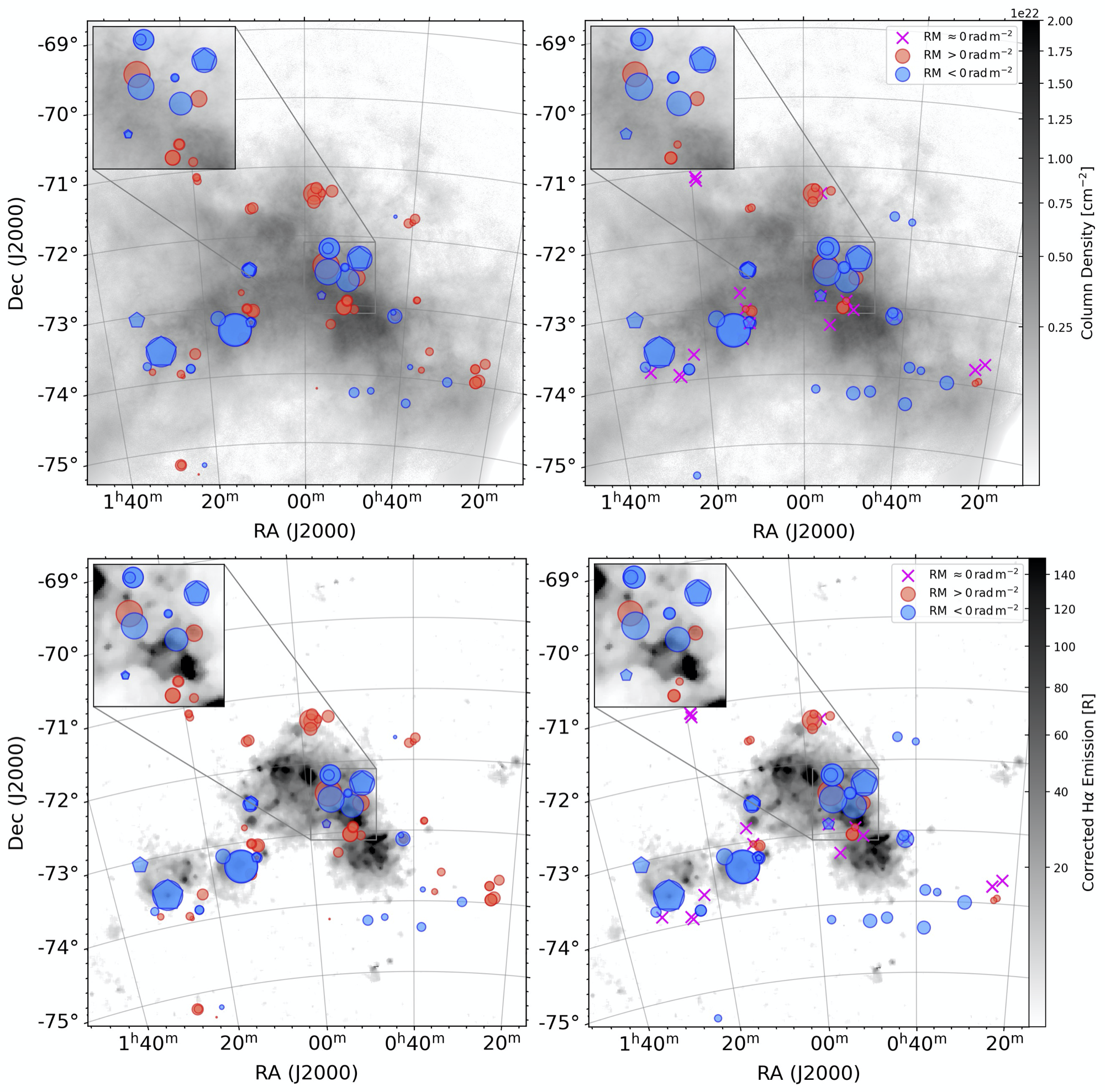}
    \caption{Plot of the RM data before (left-column) and after (right-column) foreground subtraction with $\times 2$ zoom in sections for the centre of the Bar (top left corners). The RM data presented are from this work and the RMs determined by \protect\cite{Mao2008} (represented as pentagons). The foreground model used is from \protect\cite{Mao2008}. The area of each shape is scaled by the magnitude of RM based on the scaling of marker size from \href{https://matplotlib.org/stable/api/_as_gen/matplotlib.pyplot.scatter.html}{matplotlib.pyplot.scatter}. The background image for the top row is of the $\ion{H}{i}$ column density from \protect\cite{McClureGriffiths2018} over a range of $2\times10^{20}\,\mathrm{cm^{-2}}$ to $2\times10^{22}\,\mathrm{cm^{-2}}$ with square root scaling. The background image for the bottom row is an extinction corrected H$\alpha$ map from \protect\cite{Gaustad2001} of the SMC with square root scaling ranging between 3 rayleighs to 150 rayleighs.}
    \label{fig:OppForegroundSMCpeakFD}
\end{figure*}

In Figure \ref{fig:OppForegroundSMCpeakFD} we show the spatial distribution of Faraday depths before and after foreground subtraction. The resulting foreground subtracted RMs are shown in \cref{tab:observed}. The errors of $\mathrm{RM_{SMC}}$ shown in this table come from the errors associated with RM synthesis combined in quadrature with the scatter introduced by $\sigma_{\mathrm{RM,EG}}$, $\sigma_{\mathrm{RM,MW}}$, and the error in the foreground calculation shown in \cref{eqn:fore}.

\section{Results}
\label{sec:results}
Compared to previous studies of the SMC, our study is 5 times more sensitive and markedly improves the frequency range available for observation. This leads to an order of magnitude increase in the number of sources found and in the precision of RM, due to the relationship between $\lambda^2$ coverage and resolution in $F(\phi)$. 

\begin{figure}
    \centering
    \includegraphics[width=\columnwidth]{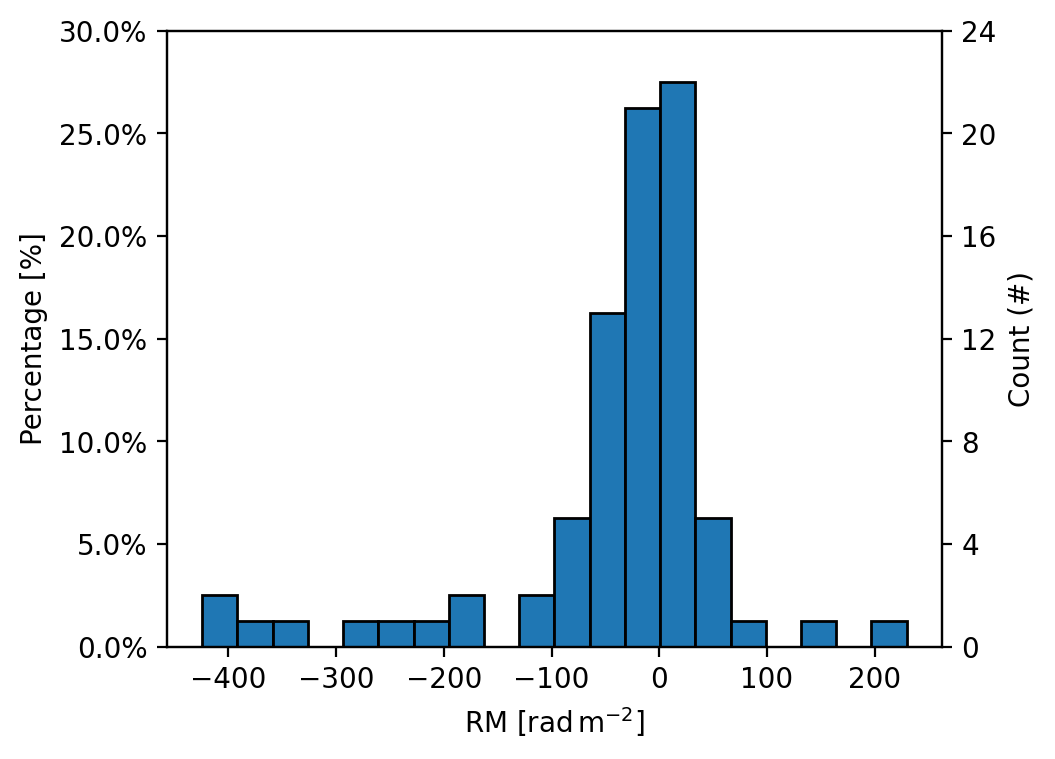}
    \caption{Distribution of RMs after Milky Way foreground subtraction. These data are placed into 20 equally sized bins.}
    \label{fig:RMdist}
\end{figure}

From the 22 observed fields, we detect 80 sources that are polarised with a minimum signal-to-noise ratio in \textit{P} of six. Of those sources, 12 were previously reported by \cite{Mao2008}. Nine of the matched sources between our study and \cite{Mao2008} were considered off-SMC in the MW foreground of \cite{Mao2008} (see section \cref{sec:foreground}) and have been excluded from our further analysis as they should be consistent with the MW foreground. From RM synthesis, after foreground subtraction, the standard deviation of Faraday depth for the 71 sources was $114_{-13}^{+17}$ rad $\mathrm{m}^{-2}$ where the upper and lower bounds are found using a bootstrap method taking the 5th and 95th percentiles. The maximum and minimum Faraday depths were $+230 \pm 23$ and $-424 \pm 8$ rad $\mathrm{m}^{-2}$. The mean and median peak Faraday depth after foreground subtraction were $-44 \pm 13$ and $-10_{-8}^{+11}$ rad $\mathrm{m}^{-2}$. The percentage of negative RMs after foreground subtraction was 59\%; this proportion is much less than that found by \cite{Mao2008} of 90\%. The distribution of the RMs after foreground subtraction is shown in \cref{fig:RMdist}.

\begin{figure}
    \centering
    \includegraphics[width=\columnwidth]{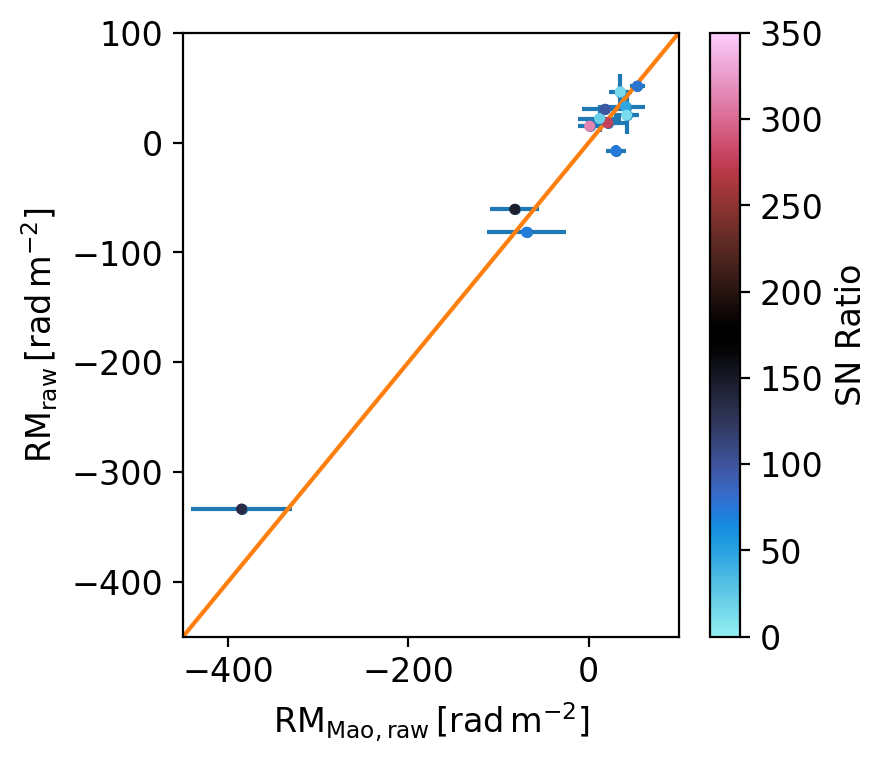}
    \caption{A comparative plot of the matched sources between our RMs and that of \protect\cite{Mao2008}. The orange line has a slope of 1. The colour of each point represents the signal-to-noise ratio of that source as calculated from our observation.}
    \label{fig:usvsmao}
\end{figure}

\cref{fig:usvsmao} shows a comparative plot of our sources that were matched to sources from \cite{Mao2008}. For all the sources in common between this study and \cite{Mao2008} except MGS2008 17, the RMs are within errors of each other as shown in \cref{tab:observed}. MGS2008 17 has a small magnitude RM, making it difficult to observe properly as instrumental effects (e.g. leakage) can produce low RM signals that are hard to disentangle from the astrophysical RM. This could explain the difference in measured RM between our observation and that of \cite{Mao2008}. 

\subsection{Faraday Complexity}
\label{sec:complexity}
As described by \cite{Alger2021}, "Faraday complexity describes whether a spectro-polarimetric observation has simple or complex magnetic structure". A Faraday spectrum may show Faraday complexity if it contains multiple Faraday depth peaks or a peak that deviates from the shape of the RMSF. We can calculate the second moment, $M_2$, of each spectrum to compare our EGSs with those of previous studies to determine how much each spectrum deviates from having a single peak with a width of the RMSF; this can act as a proxy to the Faraday complexity of a source. To calculate the $M_2$ of a spectrum, we mask all peaks that were less than seven times the noise in the cleaned $F(\phi)$ \citep[similar to the approach of][]{Anderson2015}. This signal-to-noise ratio cutoff ensures that the Faraday depths observed are physically real \citep{Hales2012,Macquart2012}. Some of the observed sources had a signal-to-noise ratio below seven as our minimum signal-to-noise ratio is six, so we set $M_2 = 0\,\mathrm{rad\,m^{-2}}$ for these sources to compare our sources with other studies of Faraday complexity \cite{Anderson2015,Livingston2021}. As stated in \cref{sec:offaxis} we have not included sources that are further than 2/3 of the primary beam FWHM away from the centre of a field. As such, we analysed the Faraday complexity of 56 sources. Faraday depths, $\phi_{i}$, were determined using the python \href{https://docs.scipy.org/doc/scipy/reference/generated/scipy.signal.find_peaks.html}{scipy.signal} package. The first moment was calculated as, 
\begin{equation}
    \langle \phi \rangle = J^{-1} \sum_{i=1}^{N} \phi_{i} |\mathrm{F}(\phi_{i})|\,(\,\mathrm{rad}\,\mathrm{m}^{-2}\,),
\end{equation}
$N$ covers all available Faraday depths. Here, the normalisation constant \textit{J} is given by,
\begin{equation}
    J = \sum_{i=1}^{N} |\mathrm{F}(\phi_{i})|\,(\,\mathrm{Jy}/\,\mathrm{beam}\,).
\end{equation}
$M_2$ was calculated as,
\begin{equation}
    M_2 = \sqrt{J^{-1} \sum_{i=1}^{N} (\phi_{i} - \langle \phi \rangle)^{2} |\mathrm{F}(\phi_{i})|}\,(\,\mathrm{rad}\,\mathrm{m}^{-2}\,). 
    \label{define:FaradayVar}
\end{equation}
\label{sec:comp}
In \cref{tab:m2} we show the mean, median, and standard deviation of $M_2$ for this work and that of \cite{Anderson2015} and \cite{Livingston2021}; \cite{Livingston2021} looked at a sample of 62 extra-galactic radio sources towards the Galactic Centre using the same telescope and frequency range as in this study and found $M_2$ for the clean components of these sources. \cite{Anderson2015} also found $M_2$ of a sample of 160 extra-galactic radio sources in a quiet patch of sky using a frequency range of 1.3--2.0 GHz, half the bandwidth used for our work. 

\begin{table}
    \centering
    \setlength{\tabcolsep}{0.3em}
    \renewcommand{\arraystretch}{1.2}
    \begin{tabular}{c c c c c}
    \hline
         Study & Mean & Median & $\sigma_{M_2}$ & $M_2 < 0.05$  \\
         & $(\mathrm{rad\,m^{-2}})$ &$(\mathrm{rad\,m^{-2}})$ & $(\mathrm{rad\,m^{-2}})$ & (\%) \\
         \hline
         \hline
         This work & 26 $\pm$ 6 & $6.1_{-0.7}^{+0.5}$& $49_{-9}^{+12} $ & 63 \\
         \cite{Anderson2015} & 6 $\pm$ 2 & 0.03 $\pm$ 0.05 & $21_{4}^{7}$ & 88 \\
         \cite{Livingston2021} & $147 \pm 20$  & $103 \pm 25 $ & $156_{-24}^{+33} $ & 5 \\
         \hline
    \end{tabular}
    \caption{Table of $M_2$ statistics for our study and those of \protect\cite{Anderson2015} and \protect\cite{Livingston2021} in units of $\mathrm{rad}\,\mathrm{m}^{-2}$. Our study and \protect\cite{Livingston2021} include those sources where $M_2 = 0\,\mathrm{rad}\,\mathrm{m}^{-2}$, whereas \protect\cite{Anderson2015} do not.}
    \label{tab:m2}
\end{table}

Sources with $M_2$ consistent with zero within errors make up 63\% of our EGSs; 88\% of the \cite{Anderson2015} sources have $M_2$ below $0.05\,\mathrm{rad}\,\mathrm{m}^{-2}$. The mean, median, and standard deviation of $M_2$ of our sample are slightly larger than \cite{Anderson2015}. As compared to our sources and the sample from \cite{Anderson2015}, 95\% of the sources from \cite{Livingston2021} have $M_2 > 0 \,\mathrm{rad\,m^{-2}}$ with a significantly higher mean, median and standard deviation $M_2$ as shown in \cref{tab:m2}. As the proportion of our EGSs with $M_2$ consistent with zero is not significantly lower than that of \cite{Anderson2015} we conclude that sight-lines that pass through the SMC do not introduce much additional complexity, unlike those of \cite{Livingston2021} going through the Galactic Centre. As a result we do not implement any further analysis to determine the possible presence of multiple RM components for each of our sources, such as QU-fitting \citep{OSullivan2012,Sun2015,Livingston2021}. 

\subsection{Spatial Trends in RM}
\label{sec:spatialRM}
\cref{fig:OppForegroundSMCpeakFD} shows the spatial distribution of RM before and after MW foreground subtraction. After foreground subtraction, there is a predominately negative RM signal (directed away from the observer). There are two regions with a higher number of large magnitude RMs which are around the centre of the Bar (RA $\sim$ $00^{\mathrm{h}}50^{\mathrm{m}}$, DEC $\sim$ $-72^{\circ}30'$) and the Wing leading towards the MB (RA $\sim$ $01^{\mathrm{h}}30^{\mathrm{m}}$, DEC $\sim$ $-73^{\circ}30'$). These two regions show increased amounts of star formation \citep{Sabbi2009} and higher H$\alpha$ emission \citep{Gaustad2001} at around 60 rayleighs (R). The region around the centre of the Bar (RA $\sim$ $00^{\mathrm{h}}50^{\mathrm{m}}$, DEC $\sim$ $-72^{\circ}30'$) also sits in a region of enhanced $\ion{H}{i}$ column density \citep{McClureGriffiths2018} at around $5\times10^{21}\mathrm{atoms\,cm^{-2}}$.

In another notable region, at the bottom of the Bar (RA $\sim$ $00^{\mathrm{h}}40^{\mathrm{m}}$, DEC $\sim$ $-74^{\circ}00'$), there is a group of negative RMs and low H$\alpha$ emission. To ensure the $\mathrm{RM_{SMC}}$ for these sight-lines is not caused by improper foreground subtraction we perform a Kolmogorov–Smirnov (K-S) test between the MW foreground at each of these source positions and the observed $\mathrm{RM_{raw}}$. We find a p-value of $0.02$, indicating that the $\mathrm{RM_{SMC}}$ of these sources is inconsistent with the MW RM foreground to a certainty of $98\%$. There is a high density of positive RMs above DEC $\sim -72^{\circ}$ and near the middle of the Bar (as shown in the zoom in region of \cref{fig:OppForegroundSMCpeakFD}) there is also a large amount of sign flipping between negative and positive RM. 

There are several points that show a nonzero RM (after MW foreground subtraction) south of the Bar, below DEC $\sim -75^{\circ}$, and above the Bar that have $\ion{H}{i}$ column densities $\lesssim 2 \times 10^{20}\,\textrm{atoms}\,\textrm{cm}^{-2}$. Lower $\ion{H}{i}$ column densities than $\sim 2 \times 10^{20}\,\textrm{atoms}\,\textrm{cm}^{-2}$ suffer from a lack of self-shielding \citep{Zheng2002}. The nonzero RM components indicate the presence of ionised gas and magnetic fields that extend out further than the self-shielded $\ion{H}{i}$, associated with the SMC. This is supported by \cite{Smart2019}, who show extended regions of H$\alpha$ emission above the Bar. 

\section{The LOS Magnetic Field Strength of the SMC}
\label{sec:discuss}
The magnetic field structure of the SMC appears to be significantly different from its closest neighbour, the LMC. The LMC has an azimuthal magnetic field \citep{Gaensler2005,Mao2012}, with a coherent strength at $1.1\mu$G and a random strength of $4.1\mu$G. \cite{Gaensler2005} found that the characteristic length scale of magneto-ionic turbulence within the LMC was 90 pc; they attribute this scale to the potential presence of evolved supernova remnants and wind bubbles that have a large impact on the morphology of ionised gas within the LMC \citep{Meaburn1980}.

The structure of the magnetic fields of the SMC and MB have been studied previously using star-light polarisation \citep{Mathewson1970,Mathewson1970b,Schmidt1970,Deinzer1973,Schmidt1976,Wayte1990,Magalhaes1990,Mao2008,Gomes2015} and RM studies \citep{Mao2008,Kaczmarek2017}. From these studies, there appears to be a connection between the magnetic field of the SMC and MB. Previous studies of the magnetic field of the SMC show a LOS magnetic field pointing away from the observer and a plane of sky field that is aligned with the MB. Additionally, they found that the total magnetic field vector of the SMC aligns with the MB \citep{Mao2008}. \cite{Kaczmarek2017} found a LOS magnetic field pointing away from us for the MB with a coherent component of strength $0.3\mu$G. The connection between the two fields could be associated with the tidally shared gas that comes off the SMC and moves towards the LMC constituting the MB. This shared field forming a `pan-Magellanic' magnetic field. 

In this section, the goal is to estimate the LOS magnetic field of the SMC ($B_{||}$) using $\mathrm{RM_{SMC}}$ combined with estimates of the electron density of the SMC. This requires us to determine a few parameters related to $n_e$ (\cref{sec:EM} and \cref{sec:DM}) that we will use in the estimation of $B_{||}$. To estimate the electron density, $n_e$, integrated along the LOS, we use models of the Dispersion Measure (DM), Emission Measure (EM), and the ionisation fraction of the SMC ($X_e$). This also requires uncertain parameters such as the filling factor, and the path length of the SMC. Finally, we discuss the possible substructures of the magnetic field of the SMC based on our estimates. 

Using \cref{eqn:phi}, we can approximate the relationship between the LOS magnetic field strength and RM to,
\begin{equation}
    B_{||} \approx \frac{\mathrm{RM}_{\mathrm{SMC}}}{\mathscr{C}\,f\langle n_{e,\mathrm{LOS}} \rangle\,L_{\mathrm{SMC}}},
    \label{eqn:MFrelationship}
\end{equation}
where $\mathscr{C}$ is the same as in \cref{eqn:RM}, $\mathrm{RM}_{\mathrm{SMC}}$ is the foreground subtracted RM of the SMC for any given sight-line, $L_{\mathrm{SMC}}$ is the SMC path length along a LOS, \textit{f} is the filling factor, and $\langle n_{e,\mathrm{LOS}} \rangle$ is the mean electron density taken along the LOS. This differentiates it from $\langle n_{e,\mathrm{SMC}} \rangle$ which is calculated for the whole SMC. 

This relationship relies on the assumption that there are no correlated fluctuations between the magnetic field and electron density  \citep{Beck2003}. The validity of this assumption in turn relies on the nature of the turbulent ISM in the SMC. If there is an anti-correlation between electron density and magnetic field fluctuations as is expected when at pressure equilibrium \citep{Beck2003}, this would mean that we would underestimate $B_{||}$; if there is a correlation, which may be the case for regions undergoing compression from supernova remnants, we would overestimate $B_{||}$. For our sight-lines that pass through star forming regions of the SMC, where supernova remnant compression may be present, we could potentially over-estimate the coherent magnetic field strength and underestimate the random magnetic field strength locally, each by a factor of 2 to 3 times. However, usually these compressive regions occupy very small sections of the total path length. Thus, their potential effect on estimating magnetic field measurements is negligible \citep{Seta2021}.

\subsection{Ionised Gas Models}
\label{sec:magmeth}
We consider six different models used to find the relationship between LOS magnetic field strength, EM, DM, filling factor, path length, $\ion{H}{i}$ column density, and RM based on the approximate relation shown in \cref{eqn:MFrelationship}. Models 1 to 3 are outlined by \cite{Mao2008}; Models 4 to 6 are outlined by \cite{Kaczmarek2017}. The models of \cite{Mao2008} use DM, whereas the models of \cite{Kaczmarek2017} had to use estimates of filling factor, ionisation fraction, and path length as they did not have sufficient DM pulsar data in the MB.
\begin{enumerate}
    \item Model 1 assumes a constant product between $\langle n_{e,\mathrm{LOS}} \rangle$ and the path length of the SMC ($L_{\mathrm{SMC}}$).  \cref{eqn:MFrelationship} becomes
    \begin{equation}
        B_{||} = \frac{\mathrm{RM}_{\mathrm{SMC}}}{\mathscr{C}\,\langle\mathrm{DM}_\mathrm{pulsar,\,SMC} \rangle}.
    \label{eqn:method1}
    \end{equation}
    \item Model 2 assumes a constant $\langle n_{e,\mathrm{LOS}} \rangle$ with a varying $L_{\mathrm{SMC}}$.  \cref{eqn:MFrelationship} becomes 
    \label{sec:magmath2}
    \begin{equation}
    B_{||} = \frac{\mathrm{RM}_{\mathrm{SMC}}}{\mathscr{C}\,\langle\mathrm{DM}_\mathrm{pulsar,\,SMC} \rangle} \frac{\mathrm{EM}_\mathrm{\,SMC}}{\mathrm{EM_{source}}}.
    \label{eqn:method2}
    \end{equation}
    Here $\mathrm{EM_{source}}$ is the EM for the LOS of the source, and $\mathrm{EM}_\mathrm{\,SMC}$ is the median EM across the SMC. Following the model of \cite{Mao2008}, we found a filling factor of $f = 0.43\pm0.16$ and a mean $L_{\mathrm{SMC}}$ of $5.8 \pm 2.2$ kpc, consistent with previous findings for a SMC neutral gas depth of 3 - 7.5 kpc \citep{Stanimirovic2004,Subramanian2009,North2010,Kapakos2011,Haschke2012}. The full determination of \textit{f} and a mean path length are in \cref{sec:model2extended}.
    \item Model 3 assumes a constant product between the filling factor (\textit{f}) and $L_{\mathrm{SMC}}$, allowing $\langle n_{e,\mathrm{LOS}} \rangle$ to vary.  \cref{eqn:MFrelationship} becomes
    \begin{equation}
    B_{||} = \frac{\mathrm{RM}_{\mathrm{SMC}}}{\mathscr{C}\,\langle\mathrm{DM}_\mathrm{pulsar,\,SMC} \rangle} \sqrt{\frac{\mathrm{EM}_\mathrm{\,SMC}}{\mathrm{EM_{source}}}}.
     \label{eqn:method3}
\end{equation}
\item Model 4 assumes a constant DM, using EM, \textit{f}, and the path length of neutral hydrogen ($L_{\ion{H}{i}}$) to determine DM,
\begin{equation}
    \mathrm{DM} = (\mathrm{EM_{source}} f L_{\ion{H}{i}})^{1/2}.
\end{equation}
This relies on the assumption that $L_{\ion{H}{ii}}= f L_{\ion{H}{i}}$. \cref{eqn:MFrelationship} becomes
\begin{equation}
    B_{||} = \frac{\mathrm{RM}_{\mathrm{SMC}}}{\mathscr{C}\,(\mathrm{EM_{source}} f L_{\ion{H}{i}})^{1/2}}.
     \label{eqn:method4}
\end{equation}
For this model we use the \textit{f} and the path length determined in Model 2.
\item Model 5 assumes a direct connection between the $\ion{H}{i}$ column density and the DM through the SMC. This requires that $n_e \sim X_e n_{\ion{H}{i}}$, where $X_e$ is the fraction of ionisation of neutral gas. Additionally, Model 5 relies on the assumption that the path length of neutral gas probes the entire LOS depth of the SMC. The relation between DM and $\ion{H}{i}$ column density is
\begin{equation}
    \mathrm{DM_{pulsar,\,SMC}} = 3.24\times10^{-19}\, X_e N_{\ion{H}{i}}.
\end{equation}
The relation from \cref{eqn:MFrelationship} becomes,
    \begin{equation}
    B_{||} = 3.24\times10^{-19} \frac{\mathrm{RM}_{\mathrm{SMC}}}{\mathscr{C}\,X_e N_{\ion{H}{i}}}.
     \label{eqn:method5}
\end{equation}
This deviates from the approach of \cite{Kaczmarek2017} as we assess $N_{\ion{H}{i}}$ for each source sight-line. To find $X_e$ of the SMC we fit across all pulsar DMs for a single $X_e$. A correlation between $\mathrm{DM_{pulsar}}$ and $N_{\ion{H}{i}}$ for the MW was found by \cite{He2013}. They found that $0.30_{-0.09}^{+0.13}\,\mathrm{DM_{pulsar,\,MW}}[\mathrm{pc\,cm^{-3}}] = N_{\ion{H}{i}} [10^{20}\times\mathrm{cm^{-2}}]$ which corresponds to an average $X_e$ for the MW of $10_{-3}^{+4}\%$. For the SMC, we find a slope of $0.15\pm0.06\,\mathrm{DM_{pulsar,\,SMC}}[\mathrm{pc\,cm^{-3}}] = N_{\ion{H}{i}} [10^{20}\times\mathrm{cm^{-2}}]$, which gives an ionisation fraction of $21_{-6}^{+16}$\%. This estimate for the ionisation fraction is comparable to that derived by \cite{Kaczmarek2017} for the Wing of the Magellanic Bridge of 29\%. \cite{Kaczmarek2017} found for the $\ion{H}{i}$-Wing and the H$\alpha$-Wing of the SMC ionisation fractions of 24\% and 21\%, respectively, based on findings from \cite{Barger2013}.

\item Model 6 assumes that instead of a mix between ionised and neutral gas, the thermal electrons form an ionised skin around the neutral gas. This ionised gas is at half the density of the neutral gas \citep{Hill2009}. This requires the gas to be at pressure equilibrium on the scale of a parsec. This assumption in combination with \cref{eqn:MFrelationship} leads to
    \begin{equation}
    B_{||} = \frac{\mathrm{RM}_{\mathrm{SMC}}N_{\ion{H}{i}}}{2 \mathscr{C}\,\mathrm{EM_{source}} f^2 L_{\ion{H}{i}}}.
     \label{eqn:method6}
\end{equation}
Similarly to Model 5, we allow $N_{\ion{H}{i}}$ to vary between each LOS.
\end{enumerate}

The results of these models are shown in Table \ref{tab:mfres} using the MW foreground subtracted RMs calculated through RM synthesis. To calculate the errors associated with each estimate we used a Monte-Carlo approach. This is required to propagate the errors of both the measured and assumed quantities as both have large uncertainties. We assumed all quantities followed a Gaussian distribution with the mean set at the best estimated value and the standard deviation set by the measured or estimated spread in the value. The value for each magnetic field estimation given in \cref{tab:mfres} is the median of 10,000 iterations and the bounds for each estimate are set at the 16th and 84th percentiles. 

\subsection{Emission Measure as a Tracer of Electron Density}
\label{sec:EM}
To estimate $B_{||}$ of the SMC, we require the electron density, $n_e$ integrated along the LOS. It is not currently possible to measure $n_e$ directly for the SMC, but with assumptions about the depth of the LOS ionised medium we can use H$\alpha$ emission as an independent tracer of $n_e^2$. This is done by calculating an implied emission measure, EM, which is a function of the square average of the electron density, $\langle n_{e,\mathrm{LOS}} \rangle^2$, along the path length of ionised gas ($L_{\ion{H}{ii}}$) and filling factor \textit{f},
\begin{equation}
    \mathrm{EM} = \int_{0}^{\infty} n_e(l)^2 \mathrm{d}l = f\langle n_{e,\mathrm{LOS}} \rangle^2 L_{\ion{H}{ii}}  \,(\mathrm{pc}\,\mathrm{cm}^{-6}).
    \label{eqn:EMne}
\end{equation}
The filling factor relates the average of the square of the electron density to the square average of the electron density,
\begin{equation}
    f = \frac{\langle n_{e,\mathrm{LOS}} \rangle^2}{\langle n_{e,\mathrm{LOS}}^2 \rangle}.
\end{equation}
We note that EM can be measured indirectly using H$\alpha$ emission, along with assumptions about the electron temperature, $T_e$, of the region. We use the relation derived by \cite{Reynolds1991} between the intrinsic $\mathrm{H\alpha}$ emission of the SMC, $I_{\mathrm{SMC}}(\mathrm{H\alpha})$ (in units of rayleighs), the electron temperature of the SMC, $T_e$, and EM,
\begin{equation}
    \mathrm{EM} \approx 2.75(T_e/10^{4} \mathrm{K})^{0.9}\,I_{\mathrm{SMC}}(\mathrm{H\alpha}).
    \label{eqn:EM}
\end{equation}
We have adopted a $T_e$ for the SMC of 14000 K as adopted by \cite{Mao2008}\footnote{Found by adding $2000$ K to the mean temperature of $\ion{H}{ii}$ of $\sim 12000$ K \citep{Dufour1977}.}. We use the SMC and MW foreground dust correction adopted by \cite{Smart2019} of $A(\mathrm{H\alpha,SMC}) = 0.24$ and $A(\mathrm{H\alpha,MW}) = 0.15$ where,
\begin{equation}
    I_{\mathrm{SMC}}(\mathrm{H\alpha}) = I_{\mathrm{obs}}(\mathrm{H\alpha})e^{A(\mathrm{H\alpha,MW})/2.5}e^{A(\mathrm{H\alpha,SMC})/2.5}.
\end{equation}
$I_{\mathrm{obs}}(\mathrm{H\alpha})$ is the H$\alpha$ emission from \protect\cite{Gaustad2001}, which is sensitive to diffuse emission. Regions with $I_{\mathrm{SMC}}(\mathrm{H\alpha}) > 50$ R were masked out to calculate the EM of the SMC ISM so the calculation was not influenced by individual $\ion{H}{ii}$ regions \citep{Mao2008}. From \cref{eqn:EM}, using $I_{\mathrm{SMC}}(\mathrm{H\alpha}) < 50$ R as a filter, the median EM across the SMC, $\mathrm{EM}_\mathrm{\,SMC}$, is $ 18_{-2}^{+5} \,\mathrm{pc\,cm^{-6}}$. Using \cref{eqn:EMne} with a path length for the SMC of $5.8 \pm 2.2$ kpc and filling factor of $0.43\pm 0.16$ we find $\langle n_{e,\mathrm{LOS}} \rangle \approx 0.037\pm0.010\,\mathrm{cm}^{-3}$. 

\subsection{Dispersion Measure as a Tracer of Electron Density}
\label{sec:DM}
The DM of a pulsar is a tracer for the integral of the LOS $n_e$ which is required to calculate $B_{||}$,
\begin{equation}
    \mathrm{DM}_{\mathrm{pulsar}} = \int_{0}^{D} n_e (l) \mathrm{d}l = f\langle n_{e,\mathrm{LOS}} \rangle D \,(\mathrm{pc}\,\mathrm{cm}^{-3}).
\end{equation}
\textit{D} is the distance to the pulsar. We obtained the DM of seven known pulsar measurements that probe the LOS DM of the SMC, given in \cref{tab:PSM}. The Galactic foreground DM contribution, $\mathrm{DM}_{\mathrm{foreground}}$, was calculated using the YMW16 free electron model \citep{Yao2017}. We assume that these pulsars are evenly distributed in the depth of the SMC, making $D \approx D_{\mathrm{SMC}}$, where $D_{\mathrm{SMC}}$ is the approximate distance to somewhere within the SMC. We follow the same approach as \cite{Mao2008} to calculate the mean $\mathrm{DM}_{\mathrm{pulsar}}$ after foreground subtraction for the SMC. As such we find $\langle\mathrm{DM}_\mathrm{pulsar,\,SMC} \rangle = 184\pm17 \,\mathrm{pc}\,\mathrm{cm}^{-3}$. 

\begin{table*}
    \centering
    \setlength{\tabcolsep}{2.0em}
    \begin{tabular}{c | c c c c c c}
    \hline
         PSR Name & RAJ2000  & DECJ2000 & DM & $\mathrm{DM}_\mathrm{foreground}$ & $\mathrm{DM}_\mathrm{pulsar,\,SMC}$ & Discovery \\
          & (h m s) & ($^{\circ}$$\,'$ $\,''$) &  ($\mathrm{pc}\,\mathrm{cm}^{-3}$) & ($\mathrm{pc}\,\mathrm{cm}^{-3}$) & ($\mathrm{pc}\,\mathrm{cm}^{-3}$) & Ref.\\
         \hline 
         \hline
         
         J0043–73   & 00:43:25.86 & -73:11:18.6 & 115.1 $\pm$ 3.4 & 30.6 & 84.5 & 1 \\
         J0045-7042 & 00:45:25.69 & -70:42:07.1 & 70 $\pm$ 3 & 28.87 & 41.1 & 2\\
         J0045-7319 & 00:45:33.16 & -73:19:03.0 & 105.4 $\pm$ 0.7 & 30.7  & 74.7 & 4 \\
         J0052-72   & 00:52:28.65 & -72:05:13.5 & 158.6 $\pm$ 1.6 & 29.8 & 128.8 & 1 \\
         J0111-7131 & 01:11:28.77 & -71:31:46.8 & 76 $\pm$ 3 & 29.3  & 46.7 & 2\\
         J0113-7200 & 01:13:11.09 & -72:20:32.2 & 125.49 $\pm$ 0.03 & 30.0  & 95.5 & 3 \\
         J0131-7310 & 01:31:28.51 & -73:10:09.0 & 205.2 $\pm$ 0.7 & 31.0  & 174.2 & 2 \\
         
         \hline
    \end{tabular}
    \caption{Table of known radio pulsars in the SMC with DM measurements before (column 4) and after (column 6) foreground (column 5) subtraction. Discovery references: (1) \protect\cite{Titus2019}, (2) \protect\cite{Manchester2005,Manchester2006} (3) \protect\cite{Crawford2001}, (4) \protect\cite{McConnell1991}. $\mathrm{DM}_\mathrm{foreground}$ is from the YMW16 Galactic free electron model \protect\citep{Yao2017}.}
    \label{tab:PSM}
\end{table*}

We calculate the mean electron density of the SMC using the dispersion of pulsar DMs and the 1D dispersion of the pulsars spatial coordinates \citep{Manchester2006,Mao2008};
\begin{equation}
    \langle n_{e,\mathrm{SMC}} \rangle = \frac{\sigma_{\mathrm{DM}}}{\sigma_{\mathrm{spatial,1D}}}.
\end{equation}
To calculate $\sigma_{\mathrm{spatial,1D}}$ we assume all pulsars are within the plane of the SMC at a mean distance of 63 kpc \citep{Cioni2000}. The dispersion of pulsar DMs is $\sigma_{\mathrm{DM}} = 43 \pm 11\,\mathrm{pc}\,\mathrm{cm}^{-3}$ after foreground subtraction. The one-dimensional spatial dispersion of the pulsar positions is $\sigma_{\mathrm{spatial,1D}} = 1030\pm160\,\mathrm{pc}$. This gives the mean electron density across the SMC, $\langle n_{e,\mathrm{SMC}} \rangle = 0.042 \pm 0.012\,\mathrm{cm}^{-3}$. We elect to use $\langle n_{e,\mathrm{SMC}} \rangle$ derived from DM compared to the result found in \cref{sec:EM}, as the $\langle n_{e,\mathrm{SMC}} \rangle$ derived in \cref{sec:EM} uses both the filling factor and path length of the SMC, which are not well constrained for the SMC. 

\subsection{Summary of Ionised Gas Models}
Models 2 and 6 have coherent LOS magnetic field measurements that exceed realistic values as shown in \cref{tab:mfres}, disqualifying them both from our consideration. For Model 6, these unrealistic values likely come from the inbuilt model assumption that the ionised and neutral gas are in pressure equilibrium, which is not the case in regions of star-formation. For Model 2, these unrealistic values likely come from the assumption of a constant $\langle n_{e,\mathrm{LOS}} \rangle$. Model 1 is the simplest model as it only allows RM to vary. Model 5 has only one uncertain parameter with an assumed value, the ionisation fraction. Models 3 and 4 have two uncertain parameters, the filling factor and the path length, neither of which are well constrained for the SMC. In comparison, our independent estimate of the ionisation fraction of the SMC for Model 5 matches with previous findings \citep{Barger2013,Kaczmarek2017}. Models 3 and 4 also suffer from artefacts in the EM derived from \cite{Gaustad2001}. Weighing the strengths and weaknesses of the different models outlined above, we choose Model 5 for our estimate of the magnetic field. This is because it relies on the least number of uncertain parameters, our estimate for this parameter (the ionisation fraction) independently agrees with previous findings, and the $N_{\ion{H}{i}}$ of the SMC from \cite{McClureGriffiths2018} allows us to assess the ionised gas content of all of our sight-lines.

\begin{figure}
    \centering
    \includegraphics[width=\columnwidth]{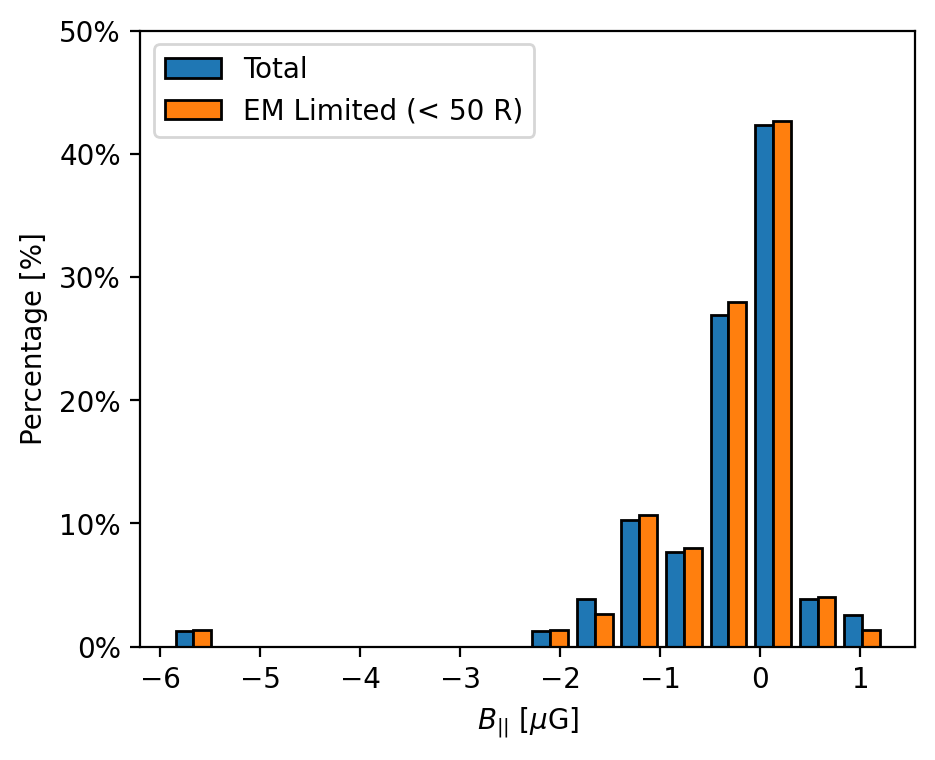}
    \caption{The distribution of magnetic strength estimations with (orange) and without (blue) a cutoff at EM$\leq 50$ R.}
    \label{fig:MFdist}
\end{figure}

We combine the 71 RMs of this survey and the 7 other RMs from \cite{Mao2008} to calculate global characteristics of the magnetic field of the SMC. As discussed in \cref{sec:results}, nine sources that were observed by both this study and by \cite{Mao2008} were used in the formulation of the MW foreground model (discussed in \cref{sec:foreground}) and as such were not included in the calculation of the magnetic fields of the SMC. The standard deviation of $B_{||}$ for the SMC using Model 5 is $0.9_{-0.2}^{+0.3}\mu$G; the mean and median coherent $B_{||}$ is $-0.3\pm0.1\,\mu$G and $-0.12\pm0.07\,\mu$G, respectively. The maximum and minimum $B_{||}$ using Model 5 is $+1.2_{-0.4}^{+1.1}\,\mu$G and $-6_{-6}^{+2}\,\mu$G. The mean $B_{||}$ of the SMC is greater than the mean of $B_{||} = -0.19\pm0.06 \mu$G found by \cite{Mao2008}. The $\langle |B_{||}| \rangle$ of the SMC is $0.52\pm0.09\,\mu$G. To separate out the effects of star formation regions from the determination of the coherent magnetic field of the SMC, we check the distribution of magnetic field estimates with and without the same EM cutoff used in \cref{sec:EM} of 50 R as shown in \cref{fig:MFdist}. There are 75 sources that have sight-lines with EM$\leq 50$ R. The mean and median $B_{||}$ is $-0.4\pm0.1\,\mu$G and $-0.14 \pm 0.07\mu$G. From \cref{fig:MFdist}, we can see a negative skew for both distributions. The negative skew of $B_{||}$ and negative mean and median $B_{||}$ is suggestive of a true (albeit weak) coherent magnetic field for the SMC. With better statistics (as discussed in \cref{sec:future}) this weak coherent field could be fully constrained and differentiated from a completely random field.

\subsection{The Random Magnetic Field}
Typically the random components of galactic magnetic fields are much stronger than the coherent components \citep{Beck2000}. We find an upper limit of the strength of the random magnetic field component of the SMC, $B_{r}$, using the method outlined by \cite{Mao2008}. This method assumes that the variation in RM is entirely from a Gaussian random magnetic field.
\begin{equation}
\label{eqn:randmag}
    B_{r} = \sqrt{\frac{3\sigma_{\mathrm{RM}}}{(\mathscr{C} n_{\mathrm{cloud}} l_0)^2 f \mathrm{N}} - 3 \langle |B_{||}| \rangle^2 (1 - f)},
\end{equation}
here $\sigma_{\mathrm{RM}}$ is the standard deviation of RM. $\mathscr{C}$ is the conversion constant from \cref{eqn:RM}. $n_{\mathrm{cloud}} = 0.098 \pm 0.016\,\mathrm{cm}^{-3}$ which is the mean ionised gas cloud electron density in the SMC \citep{Mao2008}. $l_0 = 90$ pc is the typical turbulence cell size scale along the LOS for the LMC and was used by \cite{Mao2008} for the SMC. $f = 0.43\pm0.16$ is the filling factor of the SMC which we calculated in the determination of Models 2, 3, 4, and 6. $\mathrm{N} = L/l_0$ is the number of cells along a sight-line through the SMC. We use $L = 5.8\pm2.2$ kpc (from Models 2, 3, 4, and 6) which gives us $\mathrm{N} = 64\pm24$. The $\sigma_{\mathrm{RM}}$ for the RMs of the combination of our sources and those from \cite{Mao2008} is $111\pm16\,\mathrm{rad}\,\mathrm{m}^{-2}$. Using these quantities we find a median $B_{r} = 5_{-2}^{+3}\mu$G which is a galaxy median across the SMC. The median value of $B_{r}$ is more than double that found by \cite{Mao2008}. This is to be expected: our calculated $\sigma_{\mathrm{RM}}$ and $B_{||}$ are larger than \cite{Mao2008}. From this, we observe that the random component of the magnetic field of the SMC is stronger than the ordered component of the field with a ratio of $\langle |B_{||}| \rangle/ B_{r} \sim 0.1$. This is consistent with the ratio estimated by \cite{Mao2008} of $\langle |B_{||}| \rangle/ B_{r} \sim 0.11$. We estimate that the strength of the total magnetic field of the SMC is $B_{T} = \sqrt{|B_{||}|^{2} + B_{r}^{2}} \sim 5_{-2}^{+3}\mu$G. This is consistent with the estimate of the total field strength of $\sim 5 \mu$G from \cite{Loiseau1987} and the recent estimate of $B_{T} = 5.5 \pm 1.3 \mu$G from \cite{Hassani2021}. Our estimate only serves as an upper limit as $\sigma_{\mathrm{RM}}$ may be affected by multiple external effects - such as the intrinsic scatter of RM for the background sources themselves, varying path length across SMC, varying filling factor across the SMC, and the random magnetic field of the MW foreground.

\begin{table*}
\tiny
    \centering
    \setlength{\tabcolsep}{1.1em}
    \begin{tabular}{c | c c c c c c c c c c c c}
    \hline
    Source &  Model 1 $\bar{B}_{||}$ & Error & Model 2 $\bar{B}_{||}$& Error & Model 3 $\bar{B}_{||}$ 
    & Error& Model 4 $\bar{B}_{||}$ & Error& Model 5 $\bar{B}_{||}$& Error & Model 6 $\bar{B}_{||}$& Error
    \\
    Name & ($\mu\mathrm{G}$) & ($\mu\mathrm{G}$)& ($\mu\mathrm{G}$)& ($\mu\mathrm{G}$) & ($\mu\mathrm{G}$) 
    & ($\mu\mathrm{G}$)& ($\mu\mathrm{G}$)& ($\mu\mathrm{G}$)& ($\mu\mathrm{G}$) 
    & ($\mu\mathrm{G}$)& ($\mu\mathrm{G}$) & ($\mu\mathrm{G}$) \\
    \hline
    \hline
J002248.1-734008.1&0.0&-0.1, +0.1&0.00&-0.06, +0.06&0.00&-0.08, +0.08&0.00&-0.08, +0.07&0.0&-0.4, +0.4&0.00&-0.12, +0.10\\
J002335.2-735529.1&+0.1&-0.1, +0.1&+0.06&-0.05, +0.05&+0.08&-0.06, +0.07&+0.07&-0.05, +0.07&+0.2&-0.2, +0.3&+0.1&-0.1, +0.3\\
MGS2008 42&+0.09&-0.08, +0.08&+0.04&-0.03, +0.04&+0.06&-0.05, +0.05&+0.05&-0.04, +0.05&+0.1&-0.1, +0.2&+0.1&-0.1, +0.3\\
J002412.2-735718.3&+0.09&-0.08, +0.09&+0.04&-0.03, +0.04&+0.06&-0.05, +0.06&+0.05&-0.05, +0.06&+0.1&-0.1, +0.2&+0.1&-0.1, +0.3\\
J002440.3-734542.2&0.00&-0.08, +0.08&0.00&-0.06, +0.06&0.00&-0.07, +0.06&0.00&-0.07, +0.06&0.0&-0.2, +0.2&0.0&-0.3, +0.1\\
J002440.4-734542.6&0.0&-0.1, +0.1&0.00&-0.09, +0.08&0.00&-0.09, +0.09&0.00&-0.09, +0.08&0.0&-0.3, +0.2&0.0&-0.3, +0.2\\
J003006.5-740013.3&-0.5&-0.1, +0.1&-0.6&-0.2, +0.2&-0.5&-0.2, +0.1&-0.5&-0.2, +0.1&-0.5&-0.5, +0.2&-2&-4, +1\\
MGS2008 33&0.0&-0.1, +0.1&0.00&-0.07, +0.06&0.00&-0.09, +0.08&0.00&-0.08, +0.08&0.0&-0.3, +0.2&0.0&-0.3, +0.1\\
J003545.2-735209.7&-0.1&-0.1, +0.1&-0.5&-0.4, +0.3&-0.3&-0.2, +0.2&-0.2&-0.2, +0.2&-0.2&-0.2, +0.1&-2&-4, +1\\
MGS2008 32&-0.1&-0.1, +0.1&-0.07&-0.06, +0.05&-0.09&-0.07, +0.06&-0.08&-0.07, +0.05&-0.2&-0.3, +0.1&-0.2&-0.4, +0.1\\
MGS2008 31&-0.10&-0.08, +0.07&-0.06&-0.05, +0.05&-0.08&-0.07, +0.06&-0.07&-0.07, +0.05&-0.1&-0.2, +0.1&-0.2&-0.4, +0.1\\
J003809.4-735025.0&-0.3&-0.2, +0.1&-0.2&-0.2, +0.1&-0.3&-0.2, +0.1&-0.2&-0.2, +0.1&-0.3&-0.3, +0.2&-1.0&-2.2, +0.7\\
J003824.6-742213.0&-0.4&-0.1, +0.1&-0.3&-0.1, +0.1&-0.4&-0.1, +0.1&-0.3&-0.2, +0.1&-0.7&-0.7, +0.3&-0.9&-1.8, +0.5\\
MGS2008 30&0.00&-0.08, +0.08&0.00&-0.04, +0.05&0.00&-0.06, +0.06&0.00&-0.05, +0.06&0.0&-0.5, +0.5&0.00&-0.04, +0.04\\
J004001.4-714504.6&-0.1&-0.1, +0.1&-0.10&-0.09, +0.07&-0.1&-0.1, +0.1&-0.1&-0.1, +0.1&-0.8&-1.1, +0.7&-0.06&-0.15, +0.05\\
MGS2008 29&0.00&-0.08, +0.08&0.00&-0.07, +0.07&0.00&-0.08, +0.07&0.00&-0.07, +0.07&0.0&-0.9, +0.8&0.00&-0.04, +0.04\\
J004156.4-730718.8&0.00&-0.08, +0.09&0.0&-0.2, +0.2&0.0&-0.1, +0.1&0.0&-0.1, +0.1&0.00&-0.07, +0.07&0&-1, +1\\
J004201.3-730726.9&-0.1&-0.1, +0.1&-0.2&-0.2, +0.2&-0.2&-0.1, +0.1&-0.1&-0.1, +0.1&-0.07&-0.10, +0.06&-1&-3, +1\\
J004205.9-730719.9&-0.7&-0.2, +0.1&-0.4&-0.1, +0.1&-0.5&-0.1, +0.1&-0.5&-0.2, +0.1&-0.5&-0.5, +0.2&-3&-5, +1\\
J004226.3-730418.0&-0.3&-0.1, +0.1&-0.2&-0.1, +0.1&-0.2&-0.1, +0.1&-0.2&-0.1, +0.1&-0.2&-0.2, +0.1&-1.1&-2.1, +0.6\\
J004318.3-714058.8&-0.2&-0.1, +0.1&-0.2&-0.2, +0.1&-0.2&-0.2, +0.1&-0.2&-0.2, +0.1&-1.2&-1.6, +0.8&-0.2&-0.4, +0.1\\
J004603.1-741328.6&-0.3&-0.1, +0.1&-0.3&-0.2, +0.1&-0.3&-0.1, +0.1&-0.3&-0.2, +0.1&-0.4&-0.5, +0.2&-1.0&-2.1, +0.6\\
J004934.4-721901.0&-1.7&-0.4, +0.3&-1.7&-0.5, +0.4&-1.7&-0.4, +0.3&-1.5&-0.6, +0.3&-1.8&-1.7, +0.6&-7&-13, +4\\
J004935.2-741540.8&-0.4&-0.1, +0.1&-0.4&-0.1, +0.1&-0.4&-0.1, +0.1&-0.4&-0.2, +0.1&-1.3&-1.3, +0.5&-0.6&-1.1, +0.3\\
J004957.2-723554.6&+0.5&-0.2, +0.2&+0.3&-0.2, +0.2&+0.4&-0.2, +0.2&+0.4&-0.2, +0.2&+0.3&-0.2, +0.4&+2&-1, +4\\
J005015.1-730326.4&0.0&-0.1, +0.1&0.00&-0.10, +0.10&0.0&-0.1, +0.1&0.00&-0.10, +0.09&0.00&-0.04, +0.04&0&-2, +2\\
J005140.1-723815.9&-1.5&-0.3, +0.2&-14&-22, +8&-5&-3, +1&-5&-4, +2&-0.7&-0.7, +0.2&-100&-500, +100\\
J005141.5-725603.7&+0.1&-0.1, +0.1&+0.2&-0.1, +0.1&+0.1&-0.1, +0.1&+0.1&-0.1, +0.1&+0.04&-0.03, +0.06&+2&-1, +5\\
J005141.5-725557.7&0.00&-0.08, +0.08&0.0&-0.1, +0.1&0.00&-0.09, +0.09&0.00&-0.08, +0.09&0.00&-0.03, +0.04&0&-1, +2\\
J005217.0-722703.8&-0.3&-0.1, +0.1&-0.2&-0.1, +0.1&-0.2&-0.1, +0.1&-0.2&-0.1, +0.1&-0.2&-0.2, +0.1&-0.9&-1.8, +0.6\\
J005217.5-730157.6&+0.4&-0.1, +0.1&+0.5&-0.2, +0.2&+0.4&-0.1, +0.2&+0.4&-0.1, +0.2&+0.1&-0.1, +0.1&+6&-3, +12\\
J005218.9-730153.6&+0.3&-0.1, +0.1&+0.5&-0.2, +0.2&+0.4&-0.1, +0.2&+0.4&-0.1, +0.2&+0.1&-0.1, +0.1&+5&-3, +12\\
J005218.9-722707.8&-0.4&-0.1, +0.1&-1.0&-0.5, +0.3&-0.6&-0.2, +0.2&-0.5&-0.3, +0.2&-0.3&-0.3, +0.1&-5&-11, +3\\
J005219.2-722708.8&-0.3&-0.1, +0.1&-0.9&-0.4, +0.3&-0.5&-0.2, +0.2&-0.5&-0.3, +0.2&-0.2&-0.2, +0.1&-5&-9, +3\\
J005449.7-731649.1&0.0&-0.1, +0.1&0.0&-0.2, +0.2&0.0&-0.1, +0.1&0.0&-0.1, +0.1&0.00&-0.06, +0.06&0&-2, +2\\
J005504.2-712107.8&+0.2&-0.1, +0.1&+0.3&-0.1, +0.2&+0.2&-0.1, +0.1&+0.2&-0.1, +0.1&+0.5&-0.3, +0.6&+0.4&-0.3, +0.9\\
J005522.2-721052.8&-1.2&-0.3, +0.2&-1.1&-0.3, +0.3&-1.2&-0.3, +0.2&-1.0&-0.4, +0.2&-1.1&-1.1, +0.4&-5&-10, +3\\
J005523.5-721056.8&-1.2&-0.3, +0.3&-1.1&-0.4, +0.3&-1.1&-0.3, +0.3&-1.0&-0.4, +0.3&-1.1&-1.1, +0.4&-5&-10, +3\\
J005533.2-723125.5&-1.8&-0.4, +0.3&-1.9&-0.5, +0.4&-1.9&-0.5, +0.3&-1.7&-0.7, +0.4&-0.8&-0.8, +0.3&-20&-40, +10\\
J005534.4-721056.9&-0.2&-0.1, +0.1&-0.2&-0.1, +0.1&-0.2&-0.1, +0.1&-0.2&-0.1, +0.1&-0.2&-0.2, +0.1&-1.2&-2.4, +0.7\\
J005539.9-721051.9&-0.5&-0.1, +0.1&-0.5&-0.2, +0.1&-0.5&-0.2, +0.1&-0.4&-0.2, +0.1&-0.4&-0.4, +0.2&-3&-5, +1\\
J005557.2-722605.7&+1.5&-0.3, +0.4&+1.4&-0.3, +0.4&+1.4&-0.3, +0.4&+1.3&-0.3, +0.5&+0.8&-0.3, +0.7&+12&-7, +25\\
J005652.8-712301.0&0.00&-0.08, +0.08&0.00&-0.05, +0.05&0.00&-0.06, +0.06&0.00&-0.06, +0.06&0.0&-0.2, +0.2&0.0&-0.1, +0.1\\
J005732.5-741244.0&-0.2&-0.1, +0.1&-0.2&-0.1, +0.1&-0.2&-0.1, +0.1&-0.2&-0.1, +0.1&-2&-3, +1&-0.06&-0.13, +0.04\\
J005753.8-711835.3&+0.2&-0.1, +0.1&+0.1&-0.1, +0.1&+0.1&-0.1, +0.1&+0.1&-0.1, +0.1&+0.3&-0.2, +0.4&+0.2&-0.2, +0.5\\
J005813.1-712400.8&+0.3&-0.1, +0.1&+0.3&-0.1, +0.1&+0.3&-0.1, +0.1&+0.3&-0.1, +0.1&+0.4&-0.2, +0.5&+1.0&-0.5, +1.8\\
J005817.2-712335.7&+1.0&-0.2, +0.2&+0.2&-0.1, +0.1&+0.5&-0.1, +0.1&+0.4&-0.1, +0.2&+1.2&-0.4, +1.1&+0.8&-0.4, +1.5\\
J005820.5-713040.8&+0.2&-0.2, +0.2&+1.0&-0.8, +1.1&+0.5&-0.4, +0.5&+0.4&-0.4, +0.5&+0.3&-0.3, +0.5&+3&-2, +8\\
J010931.0-713456.0&+0.1&-0.1, +0.1&+0.1&-0.1, +0.1&+0.1&-0.1, +0.1&+0.1&-0.1, +0.1&+0.3&-0.2, +0.4&+0.2&-0.1, +0.4\\
J010958.7-713543.9&+0.1&-0.1, +0.1&+0.1&-0.1, +0.1&+0.1&-0.1, +0.1&+0.1&-0.1, +0.1&+0.2&-0.2, +0.3&+0.4&-0.4, +1.1\\
J011020.4-730425.1&+0.3&-0.1, +0.1&+1.1&-0.5, +0.7&+0.6&-0.2, +0.3&+0.5&-0.2, +0.3&+0.1&-0.1, +0.1&+10&-6, +22\\
J011024.7-730450.2&+0.1&-0.1, +0.1&+0.5&-0.4, +0.5&+0.3&-0.2, +0.2&+0.3&-0.2, +0.2&+0.07&-0.05, +0.09&+5&-4, +13\\
MGS2008 18*&-0.5&-0.1, +0.1&-0.6&-0.2, +0.1&-0.6&-0.2, +0.1&-0.5&-0.2, +0.1&-0.4&-0.4, +0.2&-3&-6, +2\\
MGS2008 121*&-0.7&-0.2, +0.1&-1.8&-0.7, +0.5&-1.1&-0.3, +0.2&-1.0&-0.5, +0.3&-0.5&-0.5, +0.2&-11&-22, +6\\
J011050.0-731428.3&-0.4&-0.2, +0.2&-0.5&-0.3, +0.2&-0.4&-0.2, +0.2&-0.4&-0.3, +0.2&-0.3&-0.3, +0.2&-3&-6, +2\\
J011057.1-731406.4&-0.3&-0.1, +0.1&-0.5&-0.2, +0.2&-0.4&-0.2, +0.1&-0.4&-0.2, +0.1&-0.2&-0.2, +0.1&-4&-7, +2\\
J011130.4-730215.0&+0.1&-0.1, +0.1&+0.2&-0.2, +0.2&+0.1&-0.1, +0.1&+0.1&-0.1, +0.1&+0.06&-0.05, +0.08&+1&-1, +4\\
J011132.5-730210.0&0.00&-0.08, +0.08&0.0&-0.3, +0.3&0.0&-0.1, +0.2&0.0&-0.1, +0.1&0.00&-0.05, +0.05&0&-2, +3\\
J011226.1-732757.0&0.0&-0.1, +0.1&0.00&-0.09, +0.09&0.0&-0.1, +0.1&0.00&-0.09, +0.10&0.0&-0.1, +0.1&0.0&-0.6, +0.7\\
J011226.1-732749.0&+0.2&-0.1, +0.1&+0.1&-0.1, +0.1&+0.1&-0.1, +0.1&+0.1&-0.1, +0.1&+0.1&-0.1, +0.2&+0.7&-0.5, +1.6\\
J011227.5-724804.0&0.00&-0.10, +0.10&0.0&-0.3, +0.3&0.0&-0.2, +0.2&0.0&-0.2, +0.2&0.00&-0.07, +0.06&0&-5, +2\\
J011403.0-732007.4&-2.5&-0.5, +0.4&-2.3&-0.7, +0.5&-2.4&-0.6, +0.4&-2.2&-0.9, +0.5&-1.2&-1.2, +0.4&-20&-40, +10\\
J011408.4-732006.7&-2.8&-0.6, +0.4&-2.2&-0.6, +0.5&-2.5&-0.6, +0.4&-2.2&-0.9, +0.5&-1.3&-1.3, +0.5&-20&-40, +10\\
J011408.5-732006.2&-2.8&-0.6, +0.4&-2.1&-0.6, +0.5&-2.4&-0.6, +0.4&-2.2&-0.9, +0.5&-1.3&-1.2, +0.5&-20&-40, +10\\
J011722.1-730918.5&-0.6&-0.2, +0.1&-0.6&-0.2, +0.2&-0.6&-0.2, +0.1&-0.6&-0.2, +0.1&-0.3&-0.3, +0.1&-5&-10, +3\\
J011912.7-710833.0&0.0&-0.1, +0.1&0.0&-0.2, +0.2&0.0&-0.2, +0.2&0.0&-0.2, +0.2&0.0&-0.4, +0.4&0.0&-0.3, +0.5\\
J011917.9-710538.0&0.00&-0.08, +0.08&0.0&-0.1, +0.1&0.0&-0.1, +0.1&0.00&-0.09, +0.10&0.0&-0.3, +0.3&0.0&-0.2, +0.2\\
J011919.7-710524.0&0.00&-0.09, +0.09&0.0&-0.1, +0.1&0.0&-0.1, +0.1&0.00&-0.09, +0.11&0&-3, +4&0.00&-0.02, +0.04\\
J012235.9-733813.7&0.0&-0.2, +0.2&0.0&-0.1, +0.1&0.0&-0.1, +0.2&0.0&-0.1, +0.2&0.0&-0.2, +0.3&0.0&-0.4, +1.1\\
MGS2008 17&-0.2&-0.1, +0.1&-0.2&-0.1, +0.1&-0.2&-0.1, +0.1&-0.2&-0.1, +0.1&-1.4&-1.7, +0.8&-0.1&-0.2, +0.1\\
J012348.7-735034.0&-0.3&-0.1, +0.1&-0.3&-0.1, +0.1&-0.3&-0.1, +0.1&-0.3&-0.1, +0.1&-0.7&-0.7, +0.3&-0.6&-1.1, +0.3\\
J012350.2-735042.0&-0.3&-0.1, +0.1&-0.3&-0.1, +0.1&-0.3&-0.1, +0.1&-0.3&-0.1, +0.1&-0.8&-0.8, +0.3&-0.6&-1.1, +0.3\\
J012430.1-752243.1&-0.1&-0.1, +0.1&-0.04&-0.03, +0.03&-0.08&-0.06, +0.05&-0.07&-0.06, +0.05&-2&-2, +1&-0.01&-0.04, +0.01\\
J012536.5-735632.1&0.0&-0.1, +0.1&0.0&-0.2, +0.2&0.0&-0.1, +0.1&0.0&-0.1, +0.1&0.0&-0.3, +0.2&0.0&-0.8, +0.3\\
J012559.3-735421.3&0.0&-0.1, +0.1&0.000&-0.007, +0.007&0.00&-0.03, +0.03&0.00&-0.03, +0.03&0.0&-0.3, +0.3&0.00&-0.02, +0.02\\
MGS2008 16&0.0&-0.1, +0.1&0.00&-0.06, +0.06&0.00&-0.09, +0.10&0.00&-0.08, +0.09&0&-10, +10&0.000&-0.004, +0.008\\
MGS2008 15&+0.2&-0.1, +0.1&+0.08&-0.06, +0.07&+0.1&-0.1, +0.1&+0.1&-0.1, +0.1&+4&-4, +9&+0.01&-0.01, +0.03\\
MGS2008 134*&-2.3&-0.5, +0.4&-2.0&-0.6, +0.5&-2.2&-0.5, +0.4&-1.9&-0.8, +0.4&-6&-6, +2&-4&-7, +2\\
J013147.6-734943.3&0.00&-0.09, +0.09&0.0&-0.1, +0.1&0.00&-0.10, +0.10&0.00&-0.09, +0.09&0.0&-0.2, +0.2&0.0&-0.3, +0.3\\
J013244.0-734414.7&-0.3&-0.1, +0.1&-0.3&-0.2, +0.1&-0.3&-0.1, +0.1&-0.3&-0.2, +0.1&-0.5&-0.5, +0.2&-0.8&-1.6, +0.5\\

\hline
    \end{tabular}
    \caption{Table of LOS magnetic field measurements through the SMC using models described in \cref{sec:magmeth}; a full machine-readable table is available as part of the supplementary material provided online. Sources with an asterisk were determined by \protect\cite{Mao2008} to lie behind the SMC. Errors are calculated using a Monte-Carlo approach with 10,000 samples assuming all model quantities follow Gaussian distributions with a mean of the best estimated value and spread based on the error of the measurement; the quoted error is the range between the 16th and 84th percentiles for each set of magnetic field estimate samples.}
    \label{tab:mfres}
\end{table*}

\subsection{Features of the Magnetic Fields of the SMC and Surroundings}
\label{sec:losmf}
\begin{figure*}
    \centering
    \includegraphics[width=2\columnwidth]{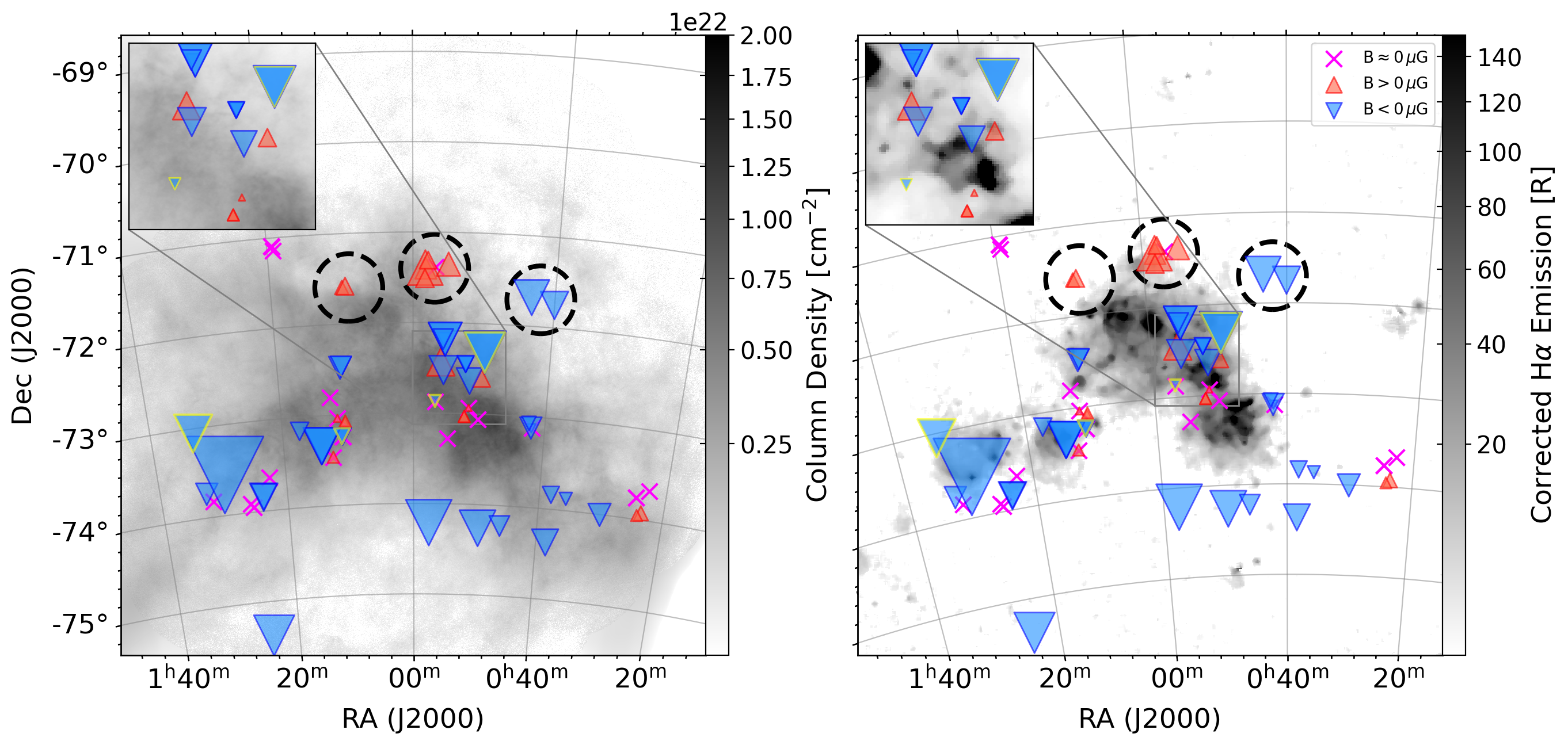}
    \caption{The 2D LOS magnetic field map of SMC using Model 5 with zoom in sections for the centre of the Bar (top left corners). Points outlined in yellow are magnetic field estimations using the RMs from \protect\cite{Mao2008}. Black dotted circles highlight three possible outflow regions further discussed in \protect\ref{sec:outflow}. The background images are the same as \cref{fig:OppForegroundSMCpeakFD}.}
    \label{fig:MFWeighted}
\end{figure*}

In \cref{fig:MFWeighted} we present the 2D LOS magnetic field of the SMC determined using Model 5, including the foreground subtracted RMs from \cite{Mao2008} in the map. Largely, we see the same trends in \cref{fig:OppForegroundSMCpeakFD} as in the 2D LOS magnetic field map. From \cref{fig:MFWeighted} we see a primarily negative $B_{||}$ field (pointed away from the observer). In the centre of the Bar there is a degree of sign flipping, indicating a large variation within the magnetic field within this region. Furthermore, unlike the magnetic fields of its closest neighbour the LMC, the SMC does not appear to have a simple magnetic field structure. The strength and general direction of the coherent magnetic field of the SMC strongly resembles that of the MB \citep{Kaczmarek2017}. Future studies into the magnetic field of the SMC and MB, discussed in \cref{sec:future}, will hopefully fill in these regions and shed more light on the complicated magnetic field structure of the Magellanic System. 

The strongest magnetic field estimation ($-6_{-6}^{+2}\,\mu$G) is at the edge of the Wing, where there is a rough line of larger magnitude negative $B_{||}$ measurements stretching from the centre of the Bar (RA = $00^{\mathrm{h}}50^{\mathrm{m}}$, DEC = $-72^{\circ}30'$) to the Wing (RA = $01^{\mathrm{h}}30^{\mathrm{m}}$, DEC = $-73^{\circ}30'$). This line points in the direction of the MB towards the LMC, but has a gap between the Bar and the Wing due to a lack of field coverage as seen in \cref{fig:field_map}. This could mean that there is some hidden structure that would be illuminated by the RMs of this region.

At the bottom of the Bar, away from strong H$\alpha$ emission, there is a group of negative magnetic field measurements, all with a similar magnitude. There is also a nonzero magnetic field measurement at DEC of $< -75^\circ$. For these regions, we expect very low densities of electrons, both because there is low EM emission and low $N_{\ion{H}{i}}$ emission. However, as noted in \cref{sec:spatialRM} there appears to be an extended H$\alpha$ emission around the SMC \citep{Smart2019}. To generate the observable RMs south of the Bar (in combination with the extended H$\alpha$ emission) requires a coherent magnetic field with an estimated mean LOS magnetic field strength of $-0.4\pm0.2\mu$G that extends out by a few degrees from the centre of the Bar. This corresponds to an on-sky distance of $\sim 2$ kpc. This is consistent with the magnetised surroundings of other local group low-mass galaxies \citep{Chyzy2011}.

\subsubsection{Magnetised Outflows}
\label{sec:outflow}
\cite{Drzazga2011} suggest that interacting galaxies could be responsible for magnetising the intergalactic medium. Interacting low-mass galaxies like the SMC could cause enough amplification to the magnetic field strength of their surroundings to be a candidate for the magnetisation of the Universe. This potential magnetisation requires a transport system that could be magnetised outflows. There are three sets of detected magnetised sight-lines, outlined with black circles in \cref{fig:MFWeighted}, that sit somewhat apart from the main body of the SMC.  All three sets are interesting because they appear to be aligned with H{\sc i} gas out-flowing from the SMC, as identified by \cite{McClureGriffiths2018}.  The regions lie close to the H{\sc i} (as shown in Figure 2 of \cite{McClureGriffiths2018}), but are offset closer to the Bar. Two of the sets, positioned to the East of the centre of the Bar, are discrepant from the overall trend of strong negative magnetic field markers near the centre of the Bar and Wing. The eastern-most set has two measurements; both positive with an average $B_{||}$ strength of $+0.3\mu$G and a difference of $0.1\mu$G. The middle set has six measurements; five of the measurements are positive and one negative. The set has an average of $+0.5\mu$G and an absolute standard deviation of $0.4\mu$G. The western-most set, is made up of two negative measurements and has an average of $-1.1\mu$G and a difference of $0.5\mu$G. This set of negative magnetic field measurements is aligned with an outflow that is more massive than the outflows aligned with the eastern and middle sets \citep{McClureGriffiths2018}. The sight-lines within each set share similar H{\sc i} emission spectra at their source positions, each with a primary peak at Local Standard of Rest velocities between $125\,\mathrm{ km\,s^{-1}}$ and $170\,\mathrm{ km\,s^{-1}}$. For the middle set with six sources, the similarity of the H{\sc i} spectra and the sign and magnitude of the individual magnetic field measurements within the set indicate that this is likely a multi-phase H{\sc i} and ionised gas outflow. For the other two sets, more Faraday rotation data is required to determine if these are associated with outflows, as discussed in \cref{sec:future}.

\cite{McClureGriffiths2018} identified regions near these three sets as spatially and kinematically distinct from the majority of the SMC H{\sc i}. Those authors argued that the H{\sc i} outflows are cold ($T < 400$ K) gas driven out of the main bar of the SMC by expanding super-bubbles associated with star formation. Based on the features' positions and velocities, they argue that they may have formed during a period of active star formation 25 -- 60 million years (Myr) ago.  The ionised gas content of the outflows is uncertain;  \cite{McClureGriffiths2018} show that there is some weak, diffuse H$\alpha$ emission in the region of the two Eastern sets in \cite{Winkler2015} and there may be some soft X-ray \citep{Sturm2014} or $\ion{O}{vi}$ emission \citep{Hoopes2002}. Clearly the detection of RMs between the main body of the SMC and these outflows implies that there must be some ionised material and magnetic field associated with these outflows, contributing to the $n_e$ required to have measurable RMs.  

The presence of these magnetised features offset from the star-forming Bar is suggestive that the magnetic field has been dragged to its position outside the galaxy by expanding super-bubbles that amplify and stretch the magnetic field lines, as observed in the MW \citep{Gao2015}, in NGC 628 \citep{Mulcahy2017}, an individual $\ion{H}{i}$ bubble in NGC 6946 \citep{Heald2012}, and in simulations  \citep{Avillez2005,Su2018}. It is also common to see the magnetisation of the surroundings of low-mass interacting galaxies \citep{Drzazga2011,Chyzy2011}. The magnetic field affecting these features could also be generated locally, but this would have to be from a yet unknown mechanism not associated with star-formation due to the apparent lack of H$\alpha$ emission. Not only do the nonzero RMs imply magnetic fields extending beyond the main body of the SMC, but they also serve as sensitive probes of ionised gas associated with the SMC outflow and hint towards a mechanism to magnetise the intergalactic medium. 

\subsubsection{Magneto-ionic Turbulence}
\label{sec:SF}
We see that for the SMC the random component of the magnetic field is stronger than the coherent, indicating it is important to consider the dynamics of the magneto-ionic environment of the SMC. The RM structure function of an RM Grid can indicate the size scales at which turbulent power is injected into the magneto-ionic medium. We follow the method outlined by \cite{Stil2011}, ensuring that each bin of our structure function contains 20 pairs of RMs. \cite{Stil2011} outline the contributions to the RM structure function at large separations as
\begin{equation}
    \mathrm{SF_{RM}} = 2\sigma_{\mathrm{int}}^2 + 2\sigma_{\mathrm{IGM}}^2 + 2\sigma_{\mathrm{ISM}}^2 + 2\sigma_{\mathrm{noise}}^2,
\end{equation}
where $\sigma_{\mathrm{int}}^2$ is the variation of RM generated in the vicinity of the AGN, $\sigma_{\mathrm{IGM}}^2$ is the contribution from the intergalactic medium, $\sigma_{\mathrm{ISM}}^2$ is the contribution of the ISM. The important contributors to $\sigma_{\mathrm{ISM}}^2$ are the MW and the SMC. $\sigma_{\mathrm{noise}}^2$ is the noise contribution of the uncertainty in measuring the RM of our sources. We expect $\sigma_{\mathrm{IGM}}^2$ to be negligible as compared to the contribution of $\sigma_{\mathrm{ISM}}^2$. To account for the variation of RM due to the MW we split $\sigma_{\mathrm{ISM}}^2$ into a contribution from the $\sigma_{\mathrm{ISM,SMC}}^2$ and $\sigma_{\mathrm{ISM,MW}}^2$. To estimate $\sigma_{\mathrm{ISM,MW}}^2$ we calculate the RM structure function of the MW RM foreground and subtract it from $\sigma_{\mathrm{ISM}}^2$ to get $\sigma_{\mathrm{ISM,SMC}}^2$. We note that the foreground subtraction of the MW contribution does not have DEC dependence. This could introduce a small North-South RM gradient into the RM structure function. 

\begin{figure}
    \centering
    \includegraphics[width=\columnwidth]{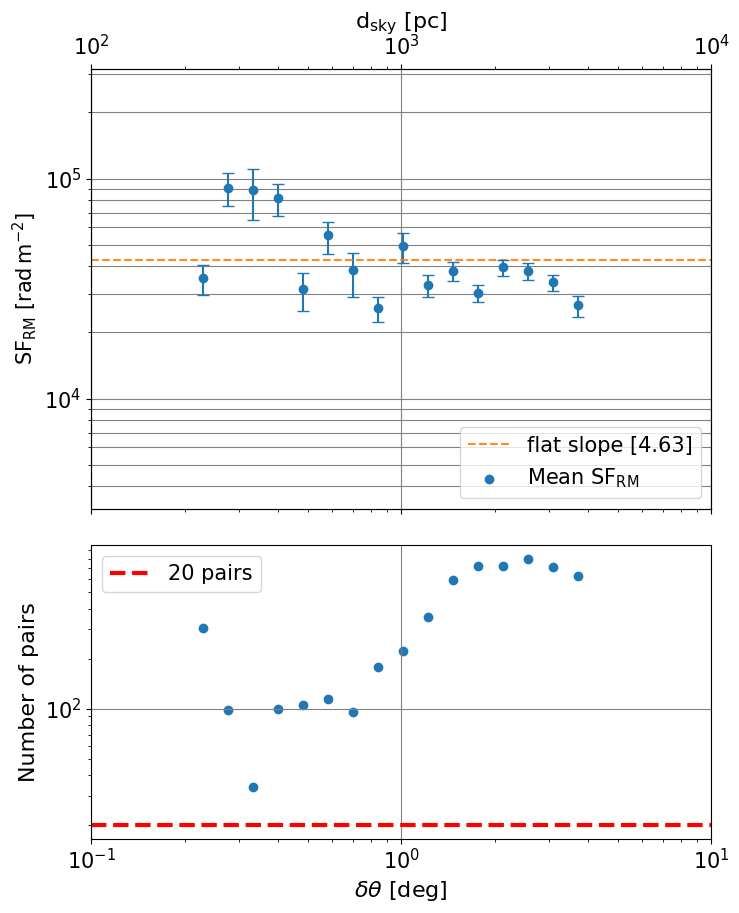}
    \caption{Plot of the RM structure function for the SMC using foreground subtracted RMs including those from \protect\cite{Mao2008} (top) and distribution of pairs of sources within bins of the RM structure function (bottom).}
    \label{fig:SMCSFSIG}
\end{figure}

To account for $2\sigma_{\mathrm{IGM}}^2 +2\sigma_{\mathrm{int}}^2$, we subtract twice the intrinsic extra-galactic RM scatter of $6\,\mathrm{rad}\,\mathrm{m}^{-2}$ found by \cite{Schnitzeler2010}. We follow the same approach as \cite{Haverkorn2004} when determining and accounting for $2\sigma_{\mathrm{noise}}^2$. We create a separate `noise' structure function which is calculated as a Gaussian with a width of $\sqrt{\mathrm{noise}^{2}}$. We create 16
bins that range over $0.2^\circ$ -- $3.7^\circ$. The smallest bin is the bottom 5th percentile of distances between sources and the largest bin is the 95th percentile of distance between sources. This corresponds to an on-sky distance range of $\sim 250 - 4100$ pc, using a distance to the SMC of 63 kpc \citep{Cioni2000}. Errors for each bin were calculated using the python module \href{https://pypi.org/project/bootstrapped/}{bootstrapped v0.0.2} with a confidence level of 95\% and 10,000 iterations to determine the spread for each bin to account for the treatment of errors in log/log plots. The bootstrapping method also accounts for bins with fewer points, giving those bins a greater spread in upper and lower bounds.  Using the RMs of this study and those of \cite{Mao2008} we determine the RM structure function as shown in \cref{fig:SMCSFSIG}.

Generally, we expect smaller scales to contribute less to the turbulent power of a medium than larger scales \citep{Kolmogorov1991,Goldreich1995}. This is called a turbulent cascade, it occurs as the energy at large scales, maintained from driving, is transferred to smaller scales where it is dissipated via viscosity. In some circumstances, turbulence is injected on small scales like in the MW arms \citep{Haverkorn2008} and the centre of the MW \citep{Livingston2021}. Observed structure functions typically follow shallow power-law slopes with a saturation point called the outer-scale, showing the scale of turbulence injection. We note that for \cref{fig:SMCSFSIG} there is no clear break in the structure function. We expect the typical size scale of coherent large-scale magnetic fields to be around 1 kpc, as is shown in the RM structure function of the M51 spiral galaxy \citep{Mao2015}. The absence of breaks in the structure function indicates the large-scale field does not vary spatially for the SMC. A spatially uniform coherent magnetic field for the SMC is consistent with the finding of \cite{Mao2008}. We have gaps within our RM Grid of the SMC, which could result in the observed lack of large-scale variation due to the coherent magnetic field of the SMC. With more complete spatial coverage (as discussed in \cref{sec:future}) the spatial variations of the coherent magnetic field could be better constrained. As we see no break in the RM structure function on small scales; the size scale of magneto-ionic turbulence must be smaller than the smallest separation we probe, which is $< 0.22^\circ$ ($< 250$ pc). 

This upper limit of $< 250$ pc is consistent with the results of \cite{Burkhart2010} in their study of the magneto-hydrodynamic (MHD) turbulence in the SMC. They found that there was no break in the spatial power spectrum for the SMC over $30\,\mathrm{pc} - 4\,\mathrm{kpc}$. Using the bispectrum\footnote{The bispectrum is the three point statistical measure that uses amplitude and phase of the correlation of signal in Fourier space.}, \cite{Burkhart2010} found a prominent break at $\sim 160$ pc. Expanding shells from supernova are attributed as the main driver of magneto-ionic turbulence throughout spiral galaxies \citep{Norman1996,Low2004,Chyzy2011} and supernova shells have sizes in the SMC between 30 pc to 800 pc \citep{Hatzidimitriou2005}. Previously, the turbulent size scale of the SMC has been assumed to be consistent with the LMC \citep{Gaensler2005,Mao2008}. The upper limit of $< 250$ pc is consistent with the turbulent size scale the LMC \citep{Gaensler2005}, attributed to the evolved supernova remnants and wind bubbles within the LMC \citep{Meaburn1980}. As the turbulent size scale of the SMC is $< 250$ pc, it is likely that expanding supernova remnant shells smaller than 250 pc are the main driver of magneto-ionic turbulence within the SMC. 

\subsection{Future Work}
\label{sec:future}
Should a larger sample of polarised extra-galactic sources behind and around the SMC become available, there will be four points to improve our understanding of the magnetic field of the SMC: 
\begin{enumerate}
    \item The first improvement would be developing a smooth foreground contribution of the MW. This would remove any ambiguities in the measurement of the LOS magnetic field of the SMC and better constrain the contribution the MW foreground has on the variation of RM across the SMC. 
    \item The second improvement would be testing to what extent the SMC magnetises its surroundings. This would test both the findings of \cite{Drzazga2011} and \cite{Chyzy2011} and investigate the magnetisation of the outflows found by \cite{McClureGriffiths2018}. Measurements from near the beginning of the Magellanic Stream would also reveal how the MW affects the magneto-ionic environment of the SMC. 
    \item The third improvement would be to refine the statistics of the magnetic field of the SMC. Allowing for smaller scales in the RM structure function to be probed up to the proposed upper limit for the turbulence cell size of 250 pc for the SMC ($\sim 0.22$ deg on-sky), any potential spatial trends of $M_2$ could be investigated (with a larger bandwidth), and the random component of the field could be properly constrained.
    \item The fourth improvement would be filling in gaps between the Bar and the Wing of the SMC and from the Wing towards the MB. This would investigate the connection between the magnetic field of the SMC and the MB, testing the `pan-Magellanic' magnetic field hypothesis.
\end{enumerate}
These four points of improvement will be achieved by upcoming broadband polarisation studies like the "Polarisation Sky Survey of the Universe's Magnetism" (POSSUM) \citep{Gaensler2010}. Such studies use the new technology of telescopes like the Australian Square Kilometre Array Pathfinder (ASKAP) \citep{Johnston2007,Hotan2021}. Points (i) and (ii) require denser RM coverage across a large angular extent (a few degrees in all directions) whereas points (iii) and (iv) require it specifically for the SMC, which may be hampered by depolarisation effects. The early science project from POSSUM of \cite{Anderson2021} achieves a source density of $\sim 25$ per square degree over $\sim 34$ square degrees. On points (i), (ii) and (iv), $\sim 34$ square degrees covers the complete extent of the SMC which would allow for both a uniform and denser RM Grid of the SMC than previously seen. On point (iii), we estimate that this source density over $\sim 34$ square degrees would correspond to a minimum RM structure function size scale (with the requirement that each bin of an RM structure function has 20 pairs) of $\sim 1.5'$ or $\sim 28$ pc.

\section{Conclusion}
\label{sec:conclude}
We have presented broadband polarisation data over 1.4 -- 3.0 GHz of 22 fields around the Small Magellanic Cloud (SMC) using the Australia Telescope Compact Array (ATCA). We found Rotation Measures (RMs) of 80 extra-galactic background sources, 71 of which were projected to be behind the SMC or its surroundings. These results were consistent with previous observations. After foreground subtraction, 59\% of RMs were negative. Notably, this is much less than the proportion of negative RMs found by \cite{Mao2008} of 90\%. We calculated the second moment, $M_2$, of all of our cleaned $F(\phi)$ and found 66\% of sources were consistent with $M_2 = 0$. This indicates that the majority of our sources were relatively simple in Faraday depth space. 

Using a variety of estimates of the line-of-sight (LOS) electron density of the SMC, we estimated $B_{||}$ for all 71 sources behind the SMC and for additional polarised sources from \cite{Mao2008}. The coherent $B_{||}$ of the SMC showed a general negative trend with a mean $B_{||} = -0.3\pm0.1\mu$G; with an ordered field of $\langle |B_{||}| \rangle = 0.52\pm0.09\mu$G. The maximum magnetic field strength found was $-6_{-6}^{+2}\mu$G, located in the Wing of the SMC. We also detect Faraday rotation in the surroundings of the SMC, indicating that the magnetic field of the SMC has a large impact on its surroundings, including three sets of magnetic field measurements coming from the top of the Bar that align with $\ion{H}{i}$ outflows \citep{McClureGriffiths2018}. The SMC appears to have a magnetic field of a typical low mass interacting galaxy, with an ordered component of $\langle B_{||} \rangle \sim -0.3\mu$G, a large random component of $\langle B_{r} \rangle \sim 5\mu$G, and turbulence likely driven by expanding supernova remnants on size scales $< 250$ pc. The primarily negative LOS magnetic of the SMC with a mean of $\langle B_{||} \rangle \sim -0.3\mu$G strongly resembles the LOS magnetic field of the MB. This observation is one further piece of evidence for a possible `pan-Magellanic' field.

\section*{Acknowledgements}
The Australia Telescope Compact Array is part of the Australia Telescope National Facility which is funded by the Australian Government for operation as a National Facility managed by CSIRO.  We acknowledge the Gomeroi people as the traditional owners of the Observatory site.  We also acknowledge the Ngunnawal, Ngunawal, and Ngambri people as the traditional owners and ongoing custodians of the land on which the Research School of Astronomy \& Astrophysics is sited at Mt Stromlo. First Nations peoples were the first astronomers of this land and make up both an important part of the history of astronomy and an integral part of astronomy going forward. 

We thank the anonymous referee for a thorough review of the work. We thank the anonymous internal referee at the Max Planck Institute for Radio Astronomy for reviewing the work. J.D.L thanks Cary Longman for helpful discussions related to the paper. This research was supported by the Australian Research Council (ARC) through grant DP160100723. J.D.L was supported by the Australian Government Research Training Program. N.M.G. acknowledges the support of the ARC through Future Fellowship FT150100024. The Dunlap Institute is funded through an endowment established by the David Dunlap family and the University of Toronto. B.M.G. acknowledges the support of the Natural Sciences and Engineering Research Council of Canada (NSERC) through grant RGPIN-2015-05948, and of the Canada Research Chairs program.

\section*{Data Availability}
The data underlying this article were accessed from the CSIRO Australia Telescope National Facility online archive at \href{https://atoa.atnf.csiro.au}{https://atoa.atnf.csiro.au}, under the project code C3086. The derived data generated in this research will be shared on reasonable request to the corresponding author.



\bibliographystyle{mnras}
\interlinepenalty=10000
\bibliography{MAIN} 



\appendix
\section{Ionised Gas Models: Model 2}
\label{sec:model2extended}
In this model we assume that there is a constant $\langle n_{e,\mathrm{LOS}} \rangle$ across the SMC and that the total path length, $L$, varies. There is observational evidence in HI velocity dispersion for varying $L$ between different lines-of-sight through the SMC \citep{Stanimirovic2004}, as well as variations in the distance modulus across the SMC \citep{2005MNRAS.359L..42L}. To calculate varying $L$, we can use DM and EM information. For this model we need to determine a physically realistic constant filling factor, \textit{f}, for the SMC, this is related to the electron density in the ionised clouds of the SMC ($n_{\mathrm{clouds}}$) and $\langle n_{e,\mathrm{LOS}} \rangle$ along a particular LOS,
\begin{equation}
    f = \langle n_{e,\mathrm{LOS}} \rangle/n_{\mathrm{clouds}}.
    \label{eqn:filling1}
\end{equation}
In the case where the electron densities in individual clouds are the same the filling factor can be expressed as,
\begin{equation}
    f = \frac{\langle n_{e,\mathrm{LOS}} \rangle^2}{\langle n_{e,\mathrm{LOS}}^2 \rangle}.
    \label{eqn:filling2}
\end{equation}
For this model we will assume that all ionised clouds have the same electron density, $n_{\mathrm{clouds}}$, and that $f$ and $\langle n_{e,\mathrm{LOS}} \rangle$ remain constant across the SMC. To determine $n_{\mathrm{clouds}}$ we can use (\ref{eqn:filling1}) and (\ref{eqn:filling2}) and the definition of $\langle\mathrm{DM}_\mathrm{\,SMC} \rangle$ to give,
\begin{equation}
    \langle\mathrm{DM}_\mathrm{\,SMC} \rangle = n_{\mathrm{clouds}} f \langle\mathrm{L}_\mathrm{\,SMC} \rangle.
    \label{eqn:cloud1}
\end{equation}
The definition of $\langle\mathrm{EM}_\mathrm{\,SMC} \rangle$ is also related to $n_{\mathrm{clouds}}$ via (\ref{eqn:filling1}) and (\ref{eqn:filling2}),
\begin{equation}
    \mathrm{EM}_\mathrm{\,SMC} = n_{\mathrm{clouds}}^2 f \langle\mathrm{L}_\mathrm{\,SMC} \rangle.
    \label{eqn:cloud2}
\end{equation}
From (\ref{eqn:cloud1}) and (\ref{eqn:cloud2}) we have that,
\begin{equation}
    n_{\mathrm{clouds}} = \frac{\mathrm{EM}_\mathrm{\,SMC}}{\langle\mathrm{DM}_\mathrm{\,SMC} \rangle}.
\end{equation}
As discussed in \cref{sec:DM}, $\langle\mathrm{DM}_\mathrm{\,SMC} \rangle = 184 \pm 17 \,\mathrm{pc}\,\mathrm{cm}^{-3}$. 
We determined $\mathrm{EM}_\mathrm{\,SMC}$ by taking the median of the foreground corrected EM calculated in \cref{sec:EM}, ignoring any values that corresponded to $I_{\mathrm{H\alpha} \mathrm{SMC}} > 50$ R. This resulted in $\mathrm{EM}_\mathrm{\,SMC} = 18_{-2}^{+5}\,\mathrm{pc}\,\mathrm{cm}^{-6}$. This gives $n_{\mathrm{clouds}} = 0.098 \pm 0.016 \,\mathrm{cm}^{-3}$. Computing \ref{eqn:filling2} with $n_{\mathrm{clouds}} = 0.098 \pm 0.016 \,\mathrm{cm}^{-3}$ and $\langle n_{e,\mathrm{SMC}} \rangle = 0.042 \pm 0.012\,\mathrm{cm}^{-3}$ (from Section 4.2), we have a filling factor of $f = 0.43\pm 0.16$ for the SMC. 

To check this model we can calculate the predicted path length (\textit{L}) through the SMC by using (\ref{eqn:cloud2}) that can be calculated for the LOS for each source;
\begin{equation}
    L_{\mathrm{source}}=\frac{\mathrm{EM_{source}}}{n_{\mathrm{clouds}}^2 f}
\end{equation}

Taking the median for all sources, we find $\langle L \rangle = 5.8\pm2.2\,\mathrm{kpc}$, which is consistent with previous findings for the SMC neutral gas depth of $3 - 7.5 $ kpc \citep{Stanimirovic2004,Subramanian2009,North2010,Kapakos2011,Haschke2012}. To include the varying $L$ information in (\ref{eqn:MFrelationship}), we can find the relative difference of each LOS by taking the ratio of $\langle\mathrm{EM}_\mathrm{\,SMC} \rangle$ and $\mathrm{EM_{source}}$,
\begin{equation}
   B_{||} = \frac{\mathrm{RM}_{\mathrm{SMC}}}{\mathscr{C}\,\langle\mathrm{DM}_\mathrm{\,SMC} \rangle} \frac{\mathrm{EM}_\mathrm{\,SMC}}{\mathrm{EM_{source}}}.
\end{equation}

\bsp	
\label{lastpage}
\end{document}